\documentclass[aps,prc,twocolumn,showpacs,amsmath,amssymb,nofootinbib]{revtex4-1}

\usepackage{graphicx,latexsym,amssymb,epsfig}
\usepackage{color}
\usepackage{times}
\usepackage{diagbox}
\usepackage{amsmath}
\usepackage[normalem]{ulem}
\usepackage{inputenc}
\usepackage{bm}
\usepackage{multirow}
\usepackage{makecell}
\usepackage{float}
\usepackage{url}
\usepackage{natbib}
\usepackage[colorlinks=true,citecolor=blue,urlcolor=blue,linkcolor=red]{hyperref}
\usepackage{lipsum}
\usepackage{tikz,xcolor,hyperref}
\usepackage{subfigure}
\usepackage{tabularx}
\usepackage[figuresright]{rotating}

\newcommand{\Msun}{\,{M}_{\odot}}
\def\fm3{\;\text{fm}^{-3}}

\definecolor{lime}{HTML}{A6CE39}
\DeclareRobustCommand{\orcidicon}{%
	\begin{tikzpicture}
	\draw[lime, fill=lime] (0,0)
	circle [radius=0.17]
	node[white] {{\fontfamily{qag}\selectfont \tiny ID}};
	\draw[white, fill=white] (-0.0625,0.095)
	circle [radius=0.008];
	\end{tikzpicture}
	\hspace{-2mm}
}
\foreach \x in {A, ..., Z}{
	\expandafter\xdef\csname orcid\x\endcsname{\noexpand\href{https://orcid.org/\csname orcidauthor\x\endcsname}{\noexpand\orcidicon}}
}

\begin{document}

\title{Impact of crust-core connection procedures on the tidal deformability of neutron stars}

\author{Junbo Pang$^{1}$}

\author{Hong Shen{\orcidA{}}$^{1}$}
\email{shennankai@gmail.com}

\author{Jinniu Hu{\orcidB{}}$^{1}$}
\email{hujinniu@nankai.edu.cn}

\affiliation{\it
$^{1}$ School of Physics, Nankai University, Tianjin 300071, China \\
}


\begin{abstract}
We study the impact of crust-core connection procedures on various neutron-star properties,
especially on the tidal deformability. We consider three types of connection procedures 
to treat the discontinuity in a nonunified equation of state around the 
crust-core transition: (1) the direct connection procedure, (2) the crossover connection 
procedure, and (3) the segmented method. Our results indicate that the mass-radius relations 
of neutron stars are almost unaffected by the details of the connection procedure. However, the tidal deformabilities of neutron stars are sensitive to 
the crust-core connection procedures. The tidal deformability is closely related to 
gravitational-wave measurements. For a canonical 1.4$M_\odot$ neutron star, uncertainties 
in the tidal deformability $\Lambda_{1.4}$ from different connection procedures can exceed 20\%.
We find that the direct connection procedure yields significantly larger
uncertainties in the tidal deformability, while the segmented method and crossover connection 
procedure provide relatively stable results.
\end{abstract}

\maketitle

\section{Introduction}
\label{sec:1}

Neutron stars are natural laboratories for studying the equation of state (EOS) of dense 
matter under extreme conditions. 
Over the past decades, astronomical observational technologies 
have achieved significant improvements, which have greatly improved our measurements and 
understanding of neutron-star properties, such as the mass, radius, and internal 
features~\cite{koeh25,chat24,asce24}. These observations provide a wealth of information 
for constraining physics of matter under extreme conditions.
Furthermore, the first direct detection of gravitational waves from a binary black 
hole merger, known as GW150914~\cite{abbo16}, launched a new era of gravitational-wave astronomy.
The observations of compact binary mergers by the LIGO and Virgo
detectors have provided valuable constraints on the neutron star EOS and its underlying theoretical framework.
It is well known that the tidal deformability of neutron stars can be inferred from 
gravitational-wave observations, although large uncertainty remains from both the measurements 
and theoretical analyses.
The milestone gravitational wave event GW170817~\cite{abbo17,abbo18} from a binary neutron-star 
merger provided an estimate of the tidal deformability and constrained the radius for canonical 
neutron stars with masses around 1.4$\Msun$~\cite{fatt18,most18,land20,chat20}.
In addition, several precise measurements for massive pulsars,
PSR J1614-2230 (1.908 $\pm$ 0.016 $M_\odot$)~\cite{arzo18},
PSR J0348+0432 (2.01  $\pm$ 0.04  $M_\odot$)~\cite{anto13}, and
PSR J0740+6620 (2.08  $\pm$ 0.07  $M_\odot$)~\cite{fons21}, 
require the predicted maximum neutron-star mass to be larger than $2\ M_\odot$,
which makes a stringent constraint on the EOS of neutron stars.
The recent observations by NICER (Neutron Star Interior Composition Explorer)
for PSR J0030+0451~\citep{mill19,rile19} and PSR J0740+6620~\citep{mill21,rile21}
provided simultaneous measurements of the mass and radius of neutron stars,
which offer further constraints on the EOS of dense matter.

The EOS plays a decisive role in studying various properties of neutron
stars~\cite{cham08,latt16,oert17}.
The EOS used as input for calculating neutron-star structure generally covers a wide density 
range, which includes three segments:
(a) the EOS of the outer crust below the neutron drip density;
(b) the EOS of the inner crust from neutron drip to crust-core transition;
(c) the EOS of the liquid core above the crust-core transition.
The outer crust extends from the surface of the star to the neutron drip density,
which is composed of spherical nuclei and a background of relativistic electron gas. 
The behavior of the outer crust is primarily determined by experimental
nuclear masses, so there are no significant differences in the EOS of the outer
crust when using different nuclear-mass models~\cite{cham08}.
As the density increases, neutrons drip out of nuclei and form a dilute neutron gas 
along with the electron gas in the inner crust.
When the density increases toward the crust-core transition, spherical nuclei may 
become unstable and nuclear shapes may vary, 
known as nuclear pasta phases~\cite{rave83,avan08,bao14a}.
The transition from the inner crust to the core, referred to as the crust-core transition, 
occurs at about 1$/$3 to 1$/$2 nuclear saturation density, depending on the nuclear models.
The uniform matter in the core includes neutrons, protons, electrons, and muons under $\beta$ equilibrium, 
extending from the crust-core transition up to a few times the nuclear saturation density.
The possible appearance of non-nucleonic degrees of freedom  
in the dense interior of neutron stars has been widely discussed in review 
papers~\cite{webe05,latt16,oert17}.
When the electron chemical potential becomes sufficiently high, negatively charged mesons such as
$\pi^{-}$ and $K^{-}$ may appear in the ground state, forming boson condensation inside neutron stars. 
The possibility of pion condensation~\cite{ohni09,khun19} 
and antikaon condensation~\cite{yue08,char14,kund23} has been investigated within various approaches. 
Other particles, such as hyperons and quarks, may appear and soften the EOS 
at higher densities~\cite{yang08,logo21,tu22,huang22,anna20}.
The appearance of these exotic components generally causes a significant reduction 
of the maximum neutron-star mass~\cite{wu19,ju21,huang22b,xia24}.
For simplicity, we do not include the non-nucleonic degrees of freedom in the present study. 

The mass and radius of a neutron star are essentially determined by the high-density core EOS, 
which should connect to the inner crust EOS at the crust-core transition. 
When the core and crust EOSs are obtained within the same nuclear model, 
the transition density can be consistently determined, and such an EOS is known as a unified EOS. There have been some studies aimed at developing a unified 
EOS~\cite{douc01,shen02,miya13,fant13,gulm15,fort16,ji19}. 
The use of a unified EOS is important for investigating the crust-core transition and detailed 
properties of neutron stars. However, most studies on neutron stars, for simplicity, employ a nonunified 
EOS and pay much more attention to the core EOS. A nonunified EOS involves a core EOS 
that is matched to a commonly used crust EOS obtained from different models~\cite{fort16,ji19}. 
Such matching may introduce a discontinuity in the nonunified EOS around 
the crust-core transition, potentially leading to uncertainties in the predictions 
of tidal deformability and other neutron-star properties. 
It has been reported in Ref.~\cite{fort16} that the crust-core connection procedure could 
slightly affect the resulting radius and crust thickness of neutron stars.
Generally, it is considered that the crust EOS has less effect on the global properties of 
neutron stars, but it may influence the resulting tidal deformabilities~\cite{ji19}.
Furthermore, the crust-core connection procedure in a nonunified EOS may also affect 
the resulting tidal deformabilities. 

Our aim in this study is to quantitatively analyze the impact of the EOS connection at the crust-core 
transition point on tidal deformability. Moreover, we intend to propose a segmented method,
which has low uncertainty in calculating neutron-star properties and helps avoid 
misleading results in studies of tidal deformability.
For this purpose, we considered three types of connection procedures 
to treat the discontinuity around the crust-core transition. 
The first one, i.e., the direct connection procedure, adopts the 
Newton polynomial interpolation method~\cite{gasc00} 
to directly connect the inner crust and core segments. 
The second one, i.e., the crossover connection procedure, employs 
a regularized calculation~\cite{alfo17}
to generate a crossover EOS between the inner crust and core segments. 
The third one, i.e., the segmented method, solves the Tolman-Oppenheimer-Volkoff (TOV) equation 
separately inside the crust and core regions, while appropriate matching conditions are imposed 
at the crust-core interface~\cite{post10}. 

To simplify the analysis of results, we adopt specific core and crust segments. 
The core EOS is obtained within the relativistic mean-field (RMF) model 
using the TM1e~\cite{ji19}, BigApple~\cite{fatt20}, and IUFSU~\cite{piek2010} 
parameterizations.
Over the past decades, the RMF approach based on various energy density functionals has been 
successfully applied to the description of finite nuclei and infinite nuclear matter~\cite{dutr14}.
The BigApple model, proposed by Fattoyev et al.~\cite{fatt20} after the discovery of GW190814,
incorporates various constraints from astrophysical observations as well as
the ground-state properties of finite nuclei.
The inner crust EOS is based on the self-consistent Thomas-Fermi approximation within the RMF model 
using the TM1e and TM1 parametrizations~\cite{bao14b,ji19}, 
which has been shown to be compatible with the constraints from astrophysical observations 
and various properties of finite nuclei~\cite{shen20}.
We note that the TM1e and TM1 models have significantly different 
behaviors of nuclear symmetry energy. The slope parameters of symmetry energy are respectively 
$L=\ $40 MeV (TM1e) and $L=\ $111 MeV (TM1). The difference in $L$ plays a key role
in the crust-core transition of neutron stars~\cite{bao15,ji19}. 
Accordingly, different combinations of the inner crust and core EOSs yield different behaviors
at the crust-core interface.
For the outer crust below the neutron drip density, we employ the well-known 
Baym-Pethick-Sutherland (BPS) EOS~\cite{baym71}, which has a relatively small 
influence on the neutron-star properties considered here.
In the present work, we focus on the uncertainty in tidal deformability induced 
by the crust-core connection procedure in a neutron star.

This article is organized as follows. 
In Sec.~\ref{sec:2}, we briefly describe the TOV equation for neutron-star calculations 
and the interpolation method for constructing the EOS.
In Sec.~\ref{sec:3}, we investigate the impact of the crust-core connection procedures 
on neutron-star properties and discuss the results, making a detailed comparison among 
the three types of connection procedures considered in this work.
Finally, a summary and conclusions are presented in Sec.~\ref{sec:4}.

\section{Formalism}
\label{sec:2}
In this section, we first describe the calculation for neutron-star properties, such as  the tidal deformability, and then describe the construction of the EOS including the
crust-core connection methods.

\subsection{Neutron-star properties}
The neutron-star properties, such as the gravitational mass ($M$) and the radius ($R$), 
can be obtained by solving the TOV 
equation~\cite{oppe39} (in units of $G=c=1$)
\begin{equation}
\begin{aligned}
\frac{\mathrm{d}P(r)}{\mathrm{d}r}=&-\frac{M(r)\varepsilon(r)}{r^2}
\left[1+\frac{P(r)}{\varepsilon(r)}\right]\\
&\times\left[1+\frac{4\pi r^3 P(r)}{M(r)}\right]
\left[1-\frac{2M(r)}{r}\right]^{-1},\\
\frac{\mathrm{d}M(r)}{\mathrm{d}r}=&\ 4\pi r^2\varepsilon(r),\\
\end{aligned}
\label{pm}
\end{equation}
where $P(r)$ and $\varepsilon(r)$ are the pressure and energy density at the radial 
coordinate $r$, respectively. $M(r)$ is the gravitational mass enclosed within the
radius $r$. The dimensionless tidal deformability $\Lambda$ of neutron stars is
expressed as~\cite{hind08}
\begin{equation}
\Lambda=\frac{2}{3}k_2C^{-5},
\label{eq:lambda}
\end{equation}
with the compactness parameter given by $C={M}/{R}$.
The second tidal Love number $k_2$ is  
calculated from~\cite{post10}
\begin{equation}
\begin{aligned}
k_2=&\left.\frac{8C^5}{5}\left(1-2C\right)^2\left[2-y_R+2C(y_R-1)\right]\right.\\
&\left.\times\left\{2C\left[6-3y_R+3C\left(5y_R-8\right)\right]\right.\right.\\
&\left.+4C^3\left[13-11y_R+C\left(3y_R-2\right)+2C^2\left(1+y_R\right)\right]\right.\\
&\left.+3\left(1-2C\right)^2\left[2-y_R+2C\left(y_R-1\right)\right]\right.\\
&\left.\times\mathrm{ln}(1-2C)\right\}^{-1},
\end{aligned}
\label{eq:k2}
\end{equation}
where $y_R=y(r)|_{r=R}$ is obtained by solving the following differential equation,
\begin{equation}
\frac{\mathrm{d}y(r)}{\mathrm{d}r}=-\frac{1}{r}\left[y(r)^2+y(r)F(r)+r^2Q(r)\right],
\label{eq:yr}
\end{equation}
with
\begin{equation}
\begin{aligned}
F(r)&=\left\{1-4\pi r^2\left[\varepsilon(r)-P(r)\right]\right\}\left[1-\frac{2M(r)}{r}\right]^{-1},\\
Q(r)&=\left\{4\pi\left[5\varepsilon(r)+9P(r)+\frac{\varepsilon(r)+P(r)}{\partial P(r)/\partial\varepsilon(r)}\right]-\frac{6}{r^2}\right\}\\
&\times\left[1-\frac{2M(r)}{r}\right]^{-1}-\left(\frac{2M(r)}{r^2}\right)^2\left[1+\frac{4\pi r^3P(r)}{M(r)}\right]^2\\
&\times\left[1-\frac{2M(r)}{r}\right]^{-2}.
\end{aligned}
\label{eq:qr}
\end{equation}

\subsection{Extract energy density from EOS}
The EOS, which is constructed under the conditions of $\beta$-equilibrium and
charge neutrality, is a critical input for solving the TOV equation.
Typically, the EOS inputted into the solving program is not a specific $P-\varepsilon$ function, 
but a set of discrete data points.
Therefore, interpolation methods are necessary to obtain the energy density
corresponding to any specific pressure value. In this work, the method we employ
is based on Newton polynomial interpolation method~\cite{gasc00}, which ensures that the interpolation results
do not deviate significantly from the polyline formed by connecting the discrete
points of the input EOS.
In this method, the energy density $\varepsilon_i$ at a given pressure $P_i$ is
calculated by the following steps:
\begin{enumerate}
\item[(1)] Locate $P_i$ within the input EOS data table, and denote the three data points 
before and after it as
\begin{equation*}
\begin{aligned}
&(P_1,\varepsilon_1),\ (P_2,\varepsilon_2),\ (P_3,\varepsilon_3),\\
&(P_4,\varepsilon_4),\ (P_5,\varepsilon_5),\ (P_6,\varepsilon_6),
\end{aligned}
\end{equation*}
where $P_i$ satisfies,
\begin{equation}
P_1<P_2<P_3<P_i<P_4<P_5<P_6.
\end{equation}
\item[(2)] Calculate four interpolation values obtained from Newton formula,
\begin{equation}
\begin{aligned}
f^{(2)}&(P_i)=\sum^4_{m=3}f[P_3,\cdots,P_m]\prod^{m-1}_{n=3}(P_i-P_n),\\
f^{(4)}_{\mathrm{L}}&(P_i)=\sum^4_{m=1}f[P_1,\cdots,P_m]\prod^{m-1}_{n=1}(P_i-P_n),\\
f^{(4)}_{\mathrm{C}}&(P_i)=\sum^5_{m=2}f[P_2,\cdots,P_m]\prod^{m-1}_{n=2}(P_i-P_n),\\
f^{(4)}_{\mathrm{R}}&(P_i)=\sum^6_{m=3}f[P_3,\cdots,P_m]\prod^{m-1}_{n=3}(P_i-P_n),
\end{aligned}
\end{equation}
where the divided difference $f[P_j,\cdots,P_k](j\le k)$ is recursively defined by the equations
\begin{equation}
\footnotesize
\begin{aligned}
f[P_j]&=\varepsilon_j,\\
f[P_j,\cdots,P_k]&=\frac{f[P_j,\cdots,P_{k-1}]-f[P_{j+1},\cdots,P_k]}{P_j-P_k}.\\
\end{aligned}
\end{equation}
The value of $f^{(2)}(P_i)$ actually employs only two data points, hence the interpolation 
polynomial is transformed into a linear function. Furthermore, we define:
\begin{equation}
\begin{aligned}
\Delta_{\mathrm{L}}&=\left|\frac{f^{(4)}_{\mathrm{L}}(P_i)-f^{(2)}(P_i)}{f^{(2)}(P_i)}\right|,\\
\Delta_{\mathrm{C}}&=\left|\frac{f^{(4)}_{\mathrm{C}}(P_i)-f^{(2)}(P_i)}{f^{(2)}(P_i)}\right|,\\
\Delta_{\mathrm{R}}&=\left|\frac{f^{(4)}_{\mathrm{R}}(P_i)-f^{(2)}(P_i)}{f^{(2)}(P_i)}\right|,
\end{aligned}
\end{equation}
to characterize the difference between the interpolation results obtained by selecting different
data points and the simple polyline of the input EOS data table.
\item[(3)] The final value of $\varepsilon_i$ is determined by the following conditions:
\end{enumerate}
\begin{itemize}
\item In the case of  $\mathrm{min}\left[\Delta_{\mathrm{L}},\Delta_{\mathrm{C}},\Delta_{\mathrm{R}}\right]>10^{-3}$, 
the value of $\varepsilon_i$ is taken directly from the linear interpolation to ensure 
that $\varepsilon_i$ does not deviate significantly from the original polyline,
\begin{equation}
\varepsilon_i=f^{(2)}(P_i).
\end{equation}
\item In the case of $\mathrm{min}\left[\Delta_{\mathrm{L}},\Delta_{\mathrm{C}},\Delta_{\mathrm{R}}\right]\le10^{-3}$, 
the value of $\varepsilon_i$ is chosen to correspond to the minimum $\Delta$,
\begin{equation}
\varepsilon_i=f^{(4)}(\mathrm{min}\left[\Delta_{\mathrm{L}},\Delta_{\mathrm{C}},
\Delta_{\mathrm{R}}\right]).
\end{equation}
\end{itemize}

 To calculate the tidal deformability $\Lambda$ by solving Eq.~(\ref{eq:yr}), 
the sound speed squared $c_s^2$ should be provided together
with the energy density $\varepsilon$ at a given pressure $P$.
Prior to solving the equations, we build an input table
that contains, for each pressure $P$, the corresponding energy density $\varepsilon$ 
and sound speed squared $c_s^2$ computed from the numerical derivative
$c_s^2=\partial P/\partial\varepsilon$.
Once the equations are solved, $c_s^2$ at any pressure $P_i$ is 
determined using Newton interpolation, in the same way as $\varepsilon_i$,
as described above.

\section{Results and Discussion}
\label{sec:3}
\begin{figure}[bp]
\centering
\includegraphics[scale=0.5]{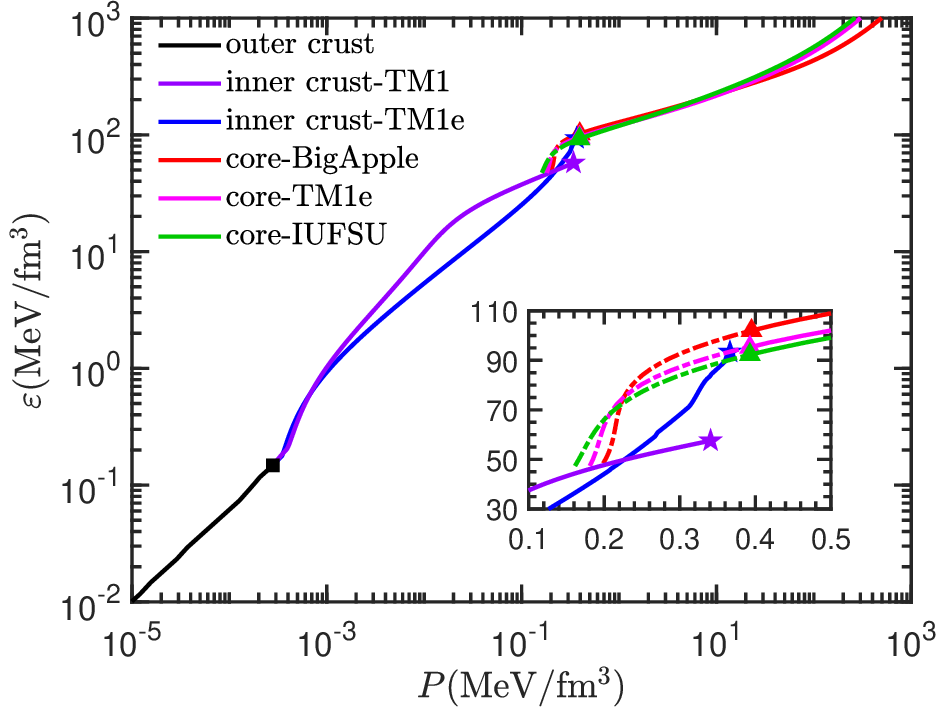}
\caption{The neutron-star matter EOSs used in this work. 
The energy density $\varepsilon$ as a function of the pressure $P$ includes three 
segments: outer crust, inner crust, and core, shown by different colors.
The BPS EOS is adopted for the outer crust, and its matching point to the
inner crust is marked by the black filled square.
The end of the inner crust is marked by the colored filled stars.
The colored filled triangles divide each core EOS into two segments. 
The connection procedure affects only the dash-dotted segments below the triangles.}
\label{eosba}
\end{figure}

It is interesting to explore how the construction of EOS influences neutron-star 
properties. In nonunified EOSs, a discontinuity in energy density typically appears at 
the crust-core transition. Such a discontinuity may induce significant uncertainties in 
the resulting tidal deformability of neutron stars. Therefore, an appropriate connection 
procedure is required to ensure a smooth EOS across the transition region, so that the tidal 
deformability can be reliably determined.
In the present work, we focus on the uncertainty in tidal deformability induced by the
connection procedure at the crust-core interface of neutron stars.

In Fig.~\ref{eosba}, we show the energy density $\varepsilon$ as a function of
the pressure $P$ including 
the crust and core segments used in the present study. 
The outer crust EOS is shown by the black solid line, and its matching point to the inner crust
is marked by the black filled square. The inner crust EOSs from different models are distinguished by color. 
The pressure at each colored filled star, denoted as $P_\star$, represents the maximum pressure at which
the nonuniform matter can exist for each model. The core EOS consists of two segments, shown by
a solid line and a dash-dotted line, joined at a colored filled triangle. 
When solving the TOV equation, the solid-line segment remains unchanged, while all connection 
procedures affect only the dash-dotted segment. 
The pressure at the colored filled triangle is denoted as $P_{\triangle}$. 
This implies that the connection procedure modifies only the core segment with $P<P_{\triangle}$.

It is noteworthy that the exact location of the crust-core transition is uncertain 
in nonunified EOSs. Therefore, we adopt different connection procedures to investigate 
the influence of the crust-core connection on the tidal deformability of neutron stars. 
In this work, we employ three types of connection procedures:
\begin{itemize}
\item Direct connection. This procedure is applied above the critical pressure $P_\star$. 
It fully retains the inner crust EOS up to $P_\star$, where the nonuniform matter vanishes, 
and then directly connects to the core EOS. The connection is implemented over a narrow 
pressure interval, with a maximum range of $[P_{\star}, P_{\triangle}]$, 
where Newton interpolation~\cite{gasc00} is employed (Sec.~\ref{sec-dc}).
\item Crossover connection. This procedure is performed in the overlapping region of 
the inner crust and core segments. We employ a regularized calculation~\cite{alfo17} to generate 
a crossover EOS between these two segments (Sec.~\ref{sec-cc}).
\item Segmented method. In this procedure, the TOV equation is solved separately inside 
the crust and core~\cite{post10}. Appropriate matching conditions are imposed at the connection pressure 
to ensure continuity and consistency between the two regions (Sec.~\ref{sec-sm}).
\end{itemize}

\subsection{Direct connection}
\label{sec-dc}

\begin{figure*}[tbp]
\centering
\begin{minipage}{0.33\textwidth}
\includegraphics[scale=0.37]{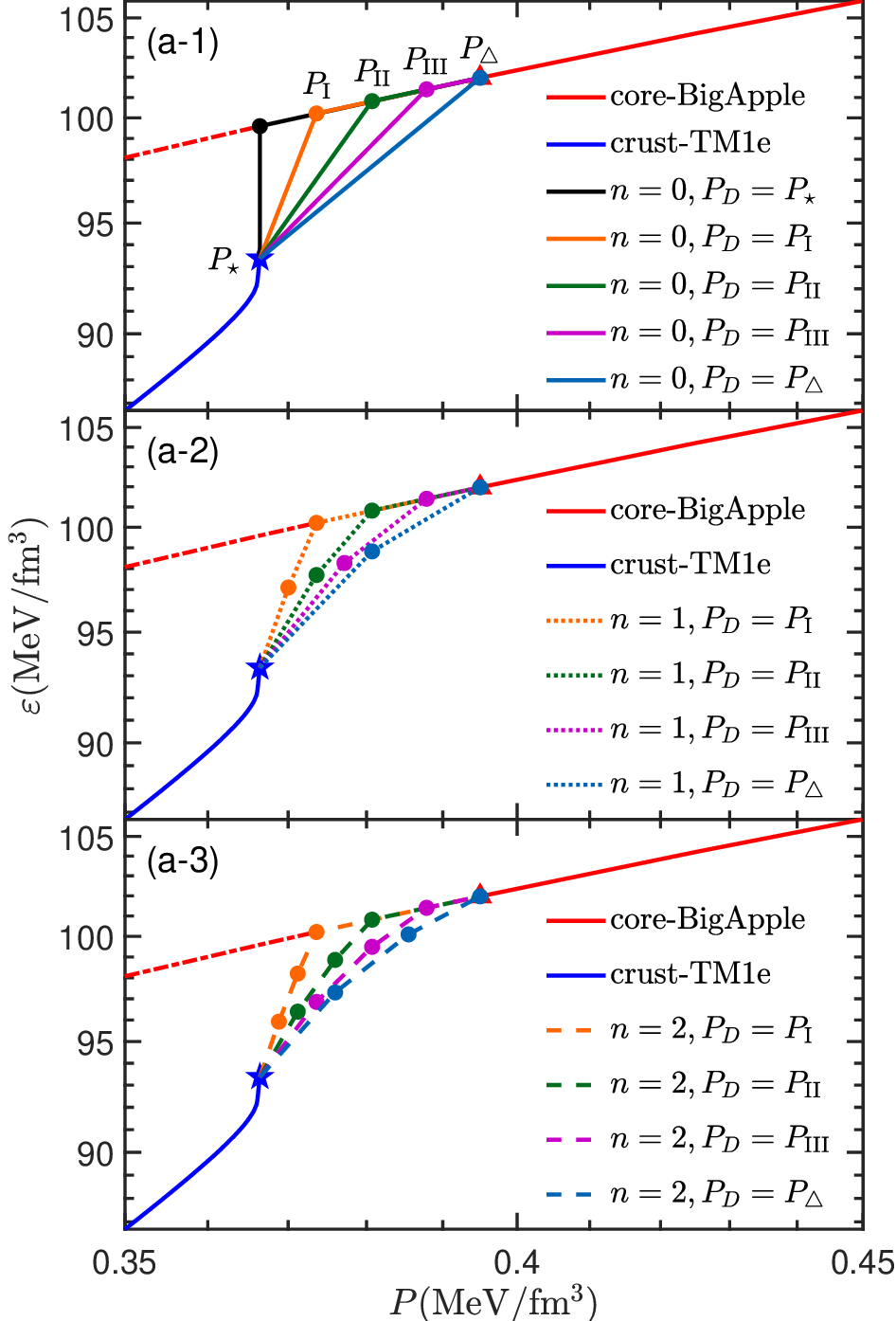}
\end{minipage}
\begin{minipage}{0.33\textwidth}
\includegraphics[scale=0.37]{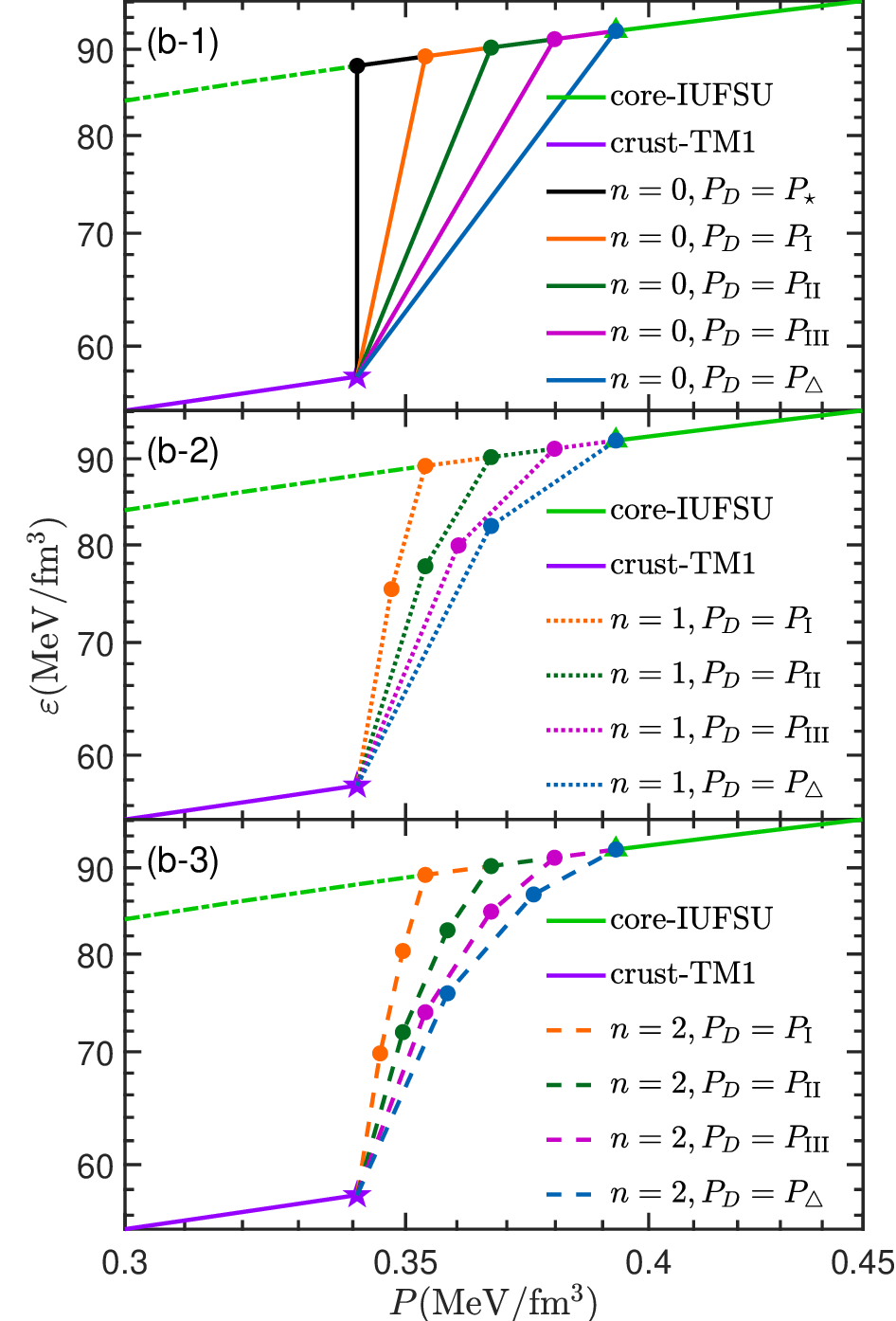}
\end{minipage}
\begin{minipage}{0.33\textwidth}
\includegraphics[scale=0.37]{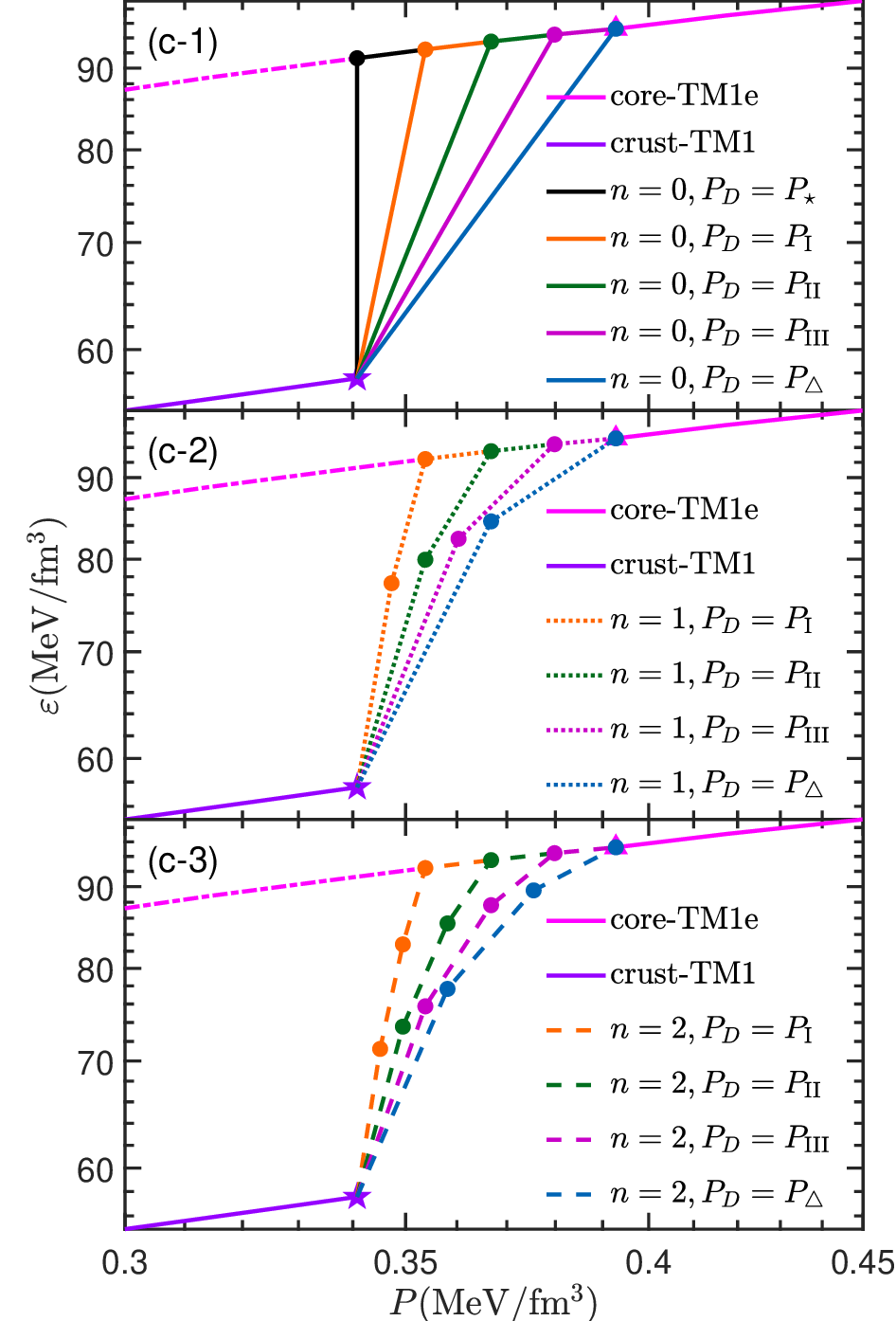}
\end{minipage}
\caption{Profiles for all nonunified EOSs constructed by the direct connection procedure. 
(a) shows the results for the combination of BigApple (core) + TM1e (crust). 
(b) is for IUFSU (core) + TM1 (crust). 
(c) is for TM1e (core) + TM1 (crust).}
\label{eosdc}
\end{figure*}
We employ the direct connection procedure~\cite{fort16} in the pressure 
interval $[P_{\star}, P_{\triangle}]$.
In Fig.~\ref{eosdc}, we present the profiles for the three sets of nonunified EOSs, constructed 
by the direct connection procedure, using different combinations of the crust and core EOSs.
All EOSs in Fig.~\ref{eosdc} share a similar structure, consisting of three segments: 
the crust EOS, the connection region, and the core EOS. 
The connection region begins at the critical pressure $P_\star$ of each inner crust EOS
and terminates at the chosen endpoint $P_D$. 
Consequently, the structure of the EOS can be represented as follows:
\begin{equation}
[\mathrm{crust}]\stackrel{(P_{\star},\varepsilon_{\star})}{|}
[\mathrm{connection}]\stackrel{(P_D,\varepsilon_D)}{|}
[\mathrm{core}],
\end{equation}
where $D$ can be $\mathrm{I}$, $\mathrm{II}$, $\mathrm{III}$, or $\triangle$. 
The pressures $P_\mathrm{I}$, $P_\mathrm{II}$ and $P_\mathrm{III}$ are the first, 
second, and third quartiles of the maximum connection interval $[P_\star, P_\triangle]$
[see Fig.~\ref{eosdc}(a-1)]. 
As mentioned above, in numerical calculations, the EOS is derived
from a polyline constructed from discrete data points.
Within the connection region, there are only a finite
number of these data points, which is denoted as $n$.
By changing $n$ and $D$, we carry out two types of connections:
\begin{enumerate}
\item[Type1.] We directly connect $(P_\star,\varepsilon_\star)$ to $(P_D,\varepsilon_D)$,
which means there is no extra data point in the connection region:
\begin{equation}
[\mathrm{crust}]| \left[(P_\star,\varepsilon_\star),(P_D,\varepsilon_D)\right]|[\mathrm{core}].
\end{equation}
In this case, $n=0$ and $D=\mathrm{I},\ \mathrm{II},\ \mathrm{III},\ \triangle$.
\item[Type2.] We add $n$ points to the connection region, where $n\ge 1$:
\begin{equation}
[\mathrm{crust}]|\left[(P_\star,\varepsilon_\star),A_1,\dots,A_n,(P_D,\varepsilon_D)\right]|
[\mathrm{core}].
\end{equation}
For $i=1,\dots,n$, $A_i$ denotes the added data point $(P_i,\varepsilon_i)$.
The pressures corresponding to these $n$ points are the equally spaced
points within the connection region, which is given by
\begin{equation}
P_i=P_{\star}+\frac{i}{n+1}\left(P_D-P_{\star}\right).
\end{equation}
The corresponding energy density $\varepsilon_i$ is obtained via four-point Newton interpolation 
based on the points $(P_\star,\varepsilon_\star)$, $(P_D,\varepsilon_D)$,
and the two closest higher-pressure data points.
\end{enumerate}
\begin{figure*}[tbp]
\centering
\begin{minipage}{0.325\textwidth}
\includegraphics[scale=0.375]{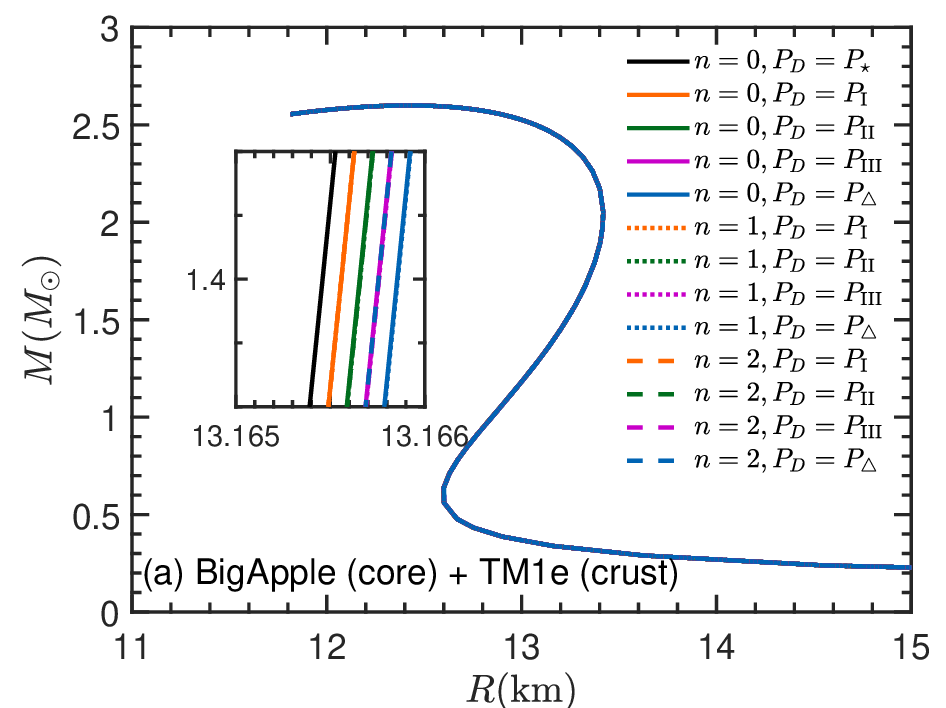}
\end{minipage}
\begin{minipage}{0.325\textwidth}
\includegraphics[scale=0.375]{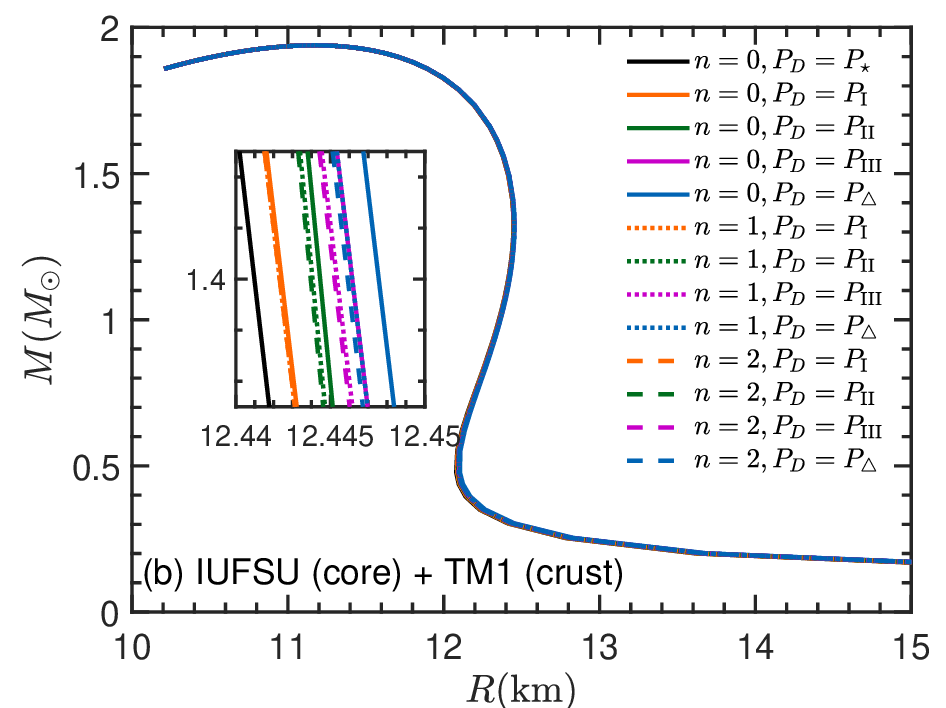}
\end{minipage}
\begin{minipage}{0.325\textwidth}
\includegraphics[scale=0.375]{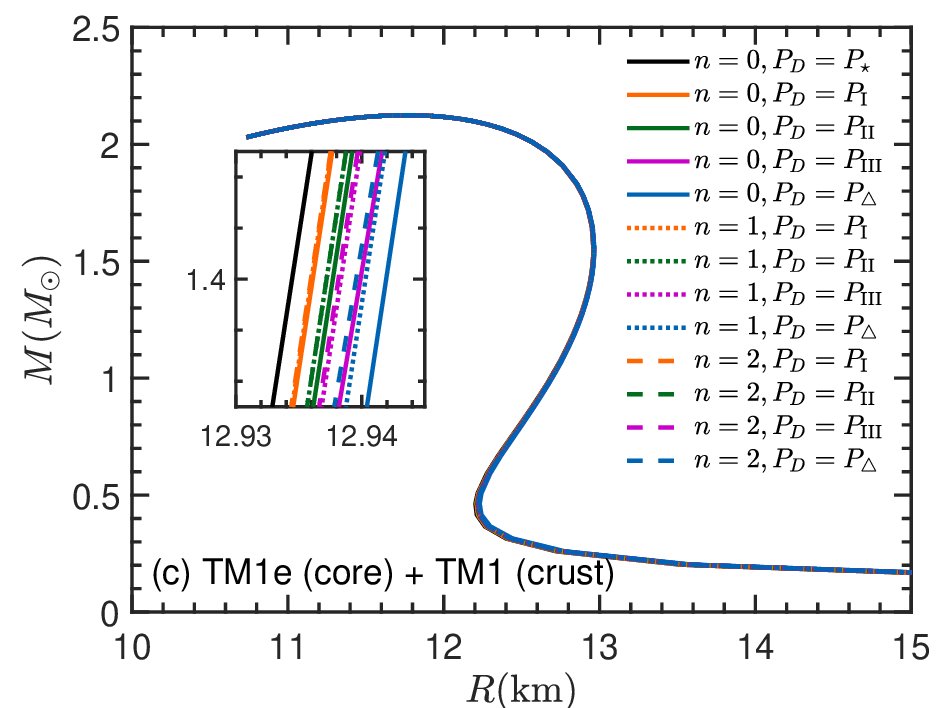}
\end{minipage}
\caption{Mass-radius relations predicted by the direct-connection EOSs 
shown in Fig.~\ref{eosdc}. The insets show more details for canonical neutron stars 
with masses around 1.4$\Msun$.}
\label{mrdc}
\end{figure*}
\begin{figure*}[tbp]
\centering
\begin{minipage}{0.325\textwidth}
\includegraphics[scale=0.37]{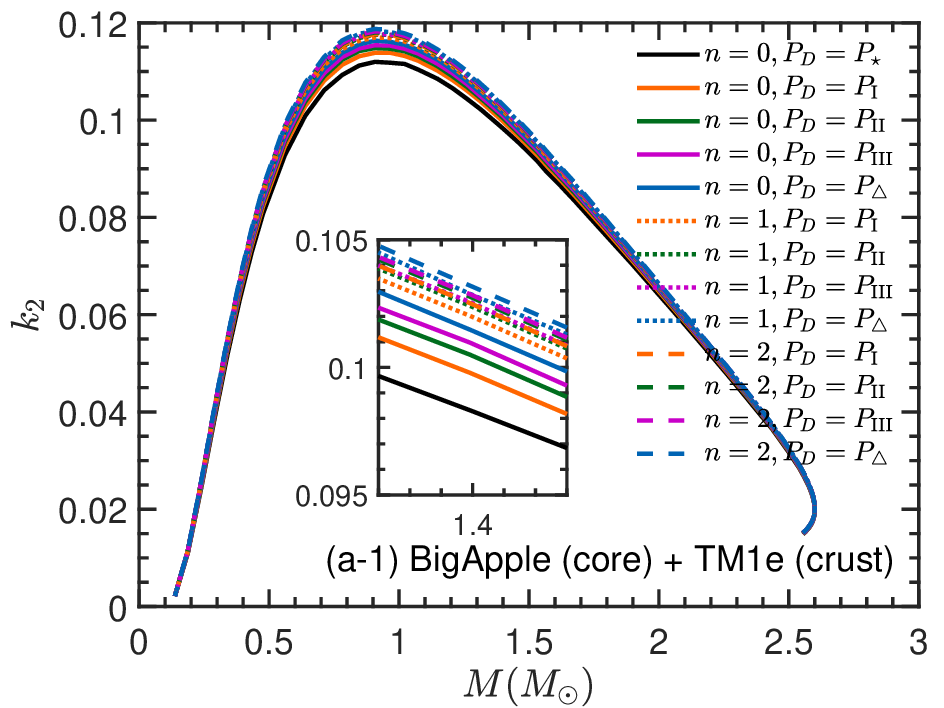}
\end{minipage}
\begin{minipage}{0.325\textwidth}
\includegraphics[scale=0.37]{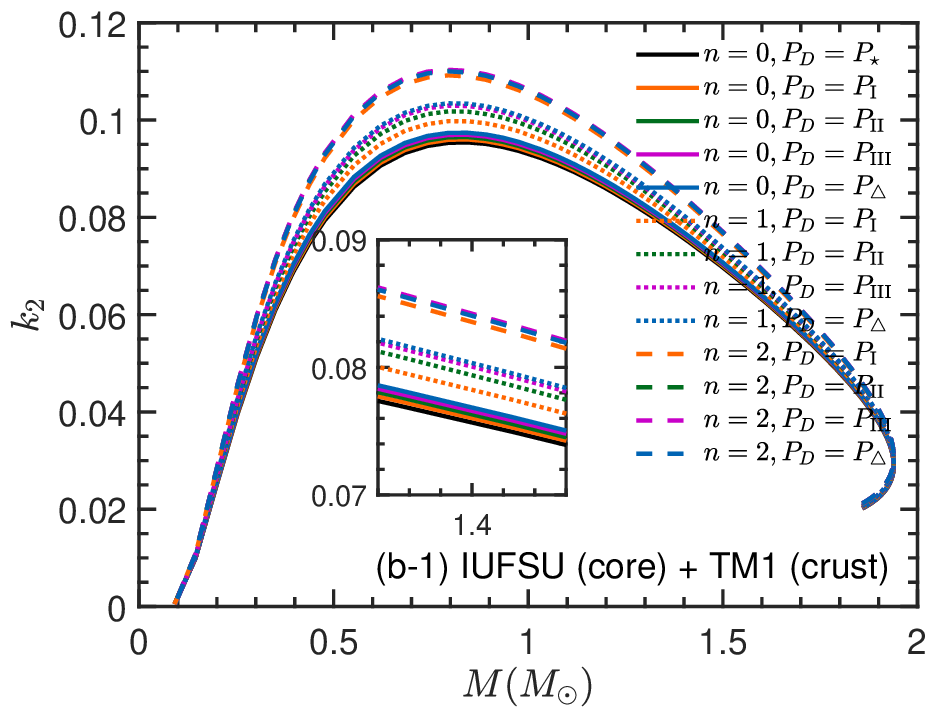}
\end{minipage}
\begin{minipage}{0.325\textwidth}
\includegraphics[scale=0.37]{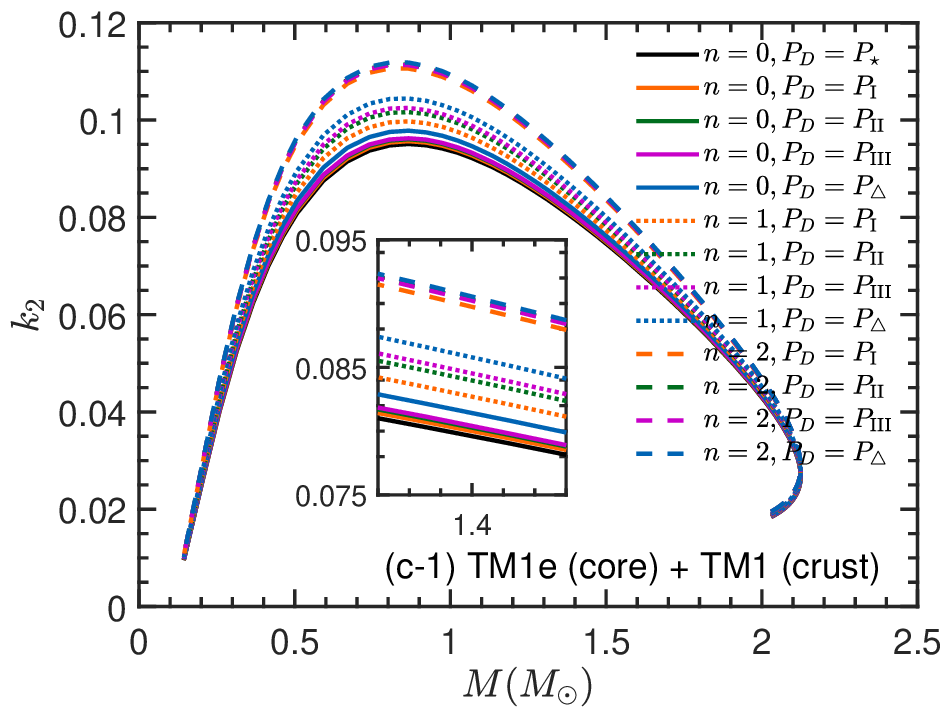}
\end{minipage}
\centering
\begin{minipage}{0.325\textwidth}
\includegraphics[scale=0.37]{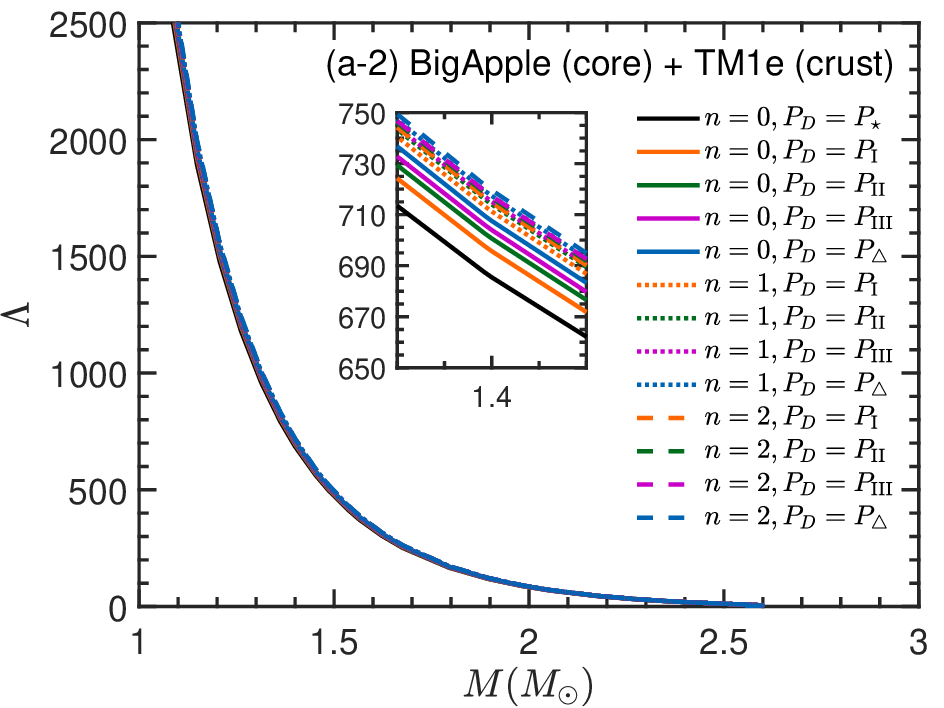}
\end{minipage}
\begin{minipage}{0.325\textwidth}
\includegraphics[scale=0.37]{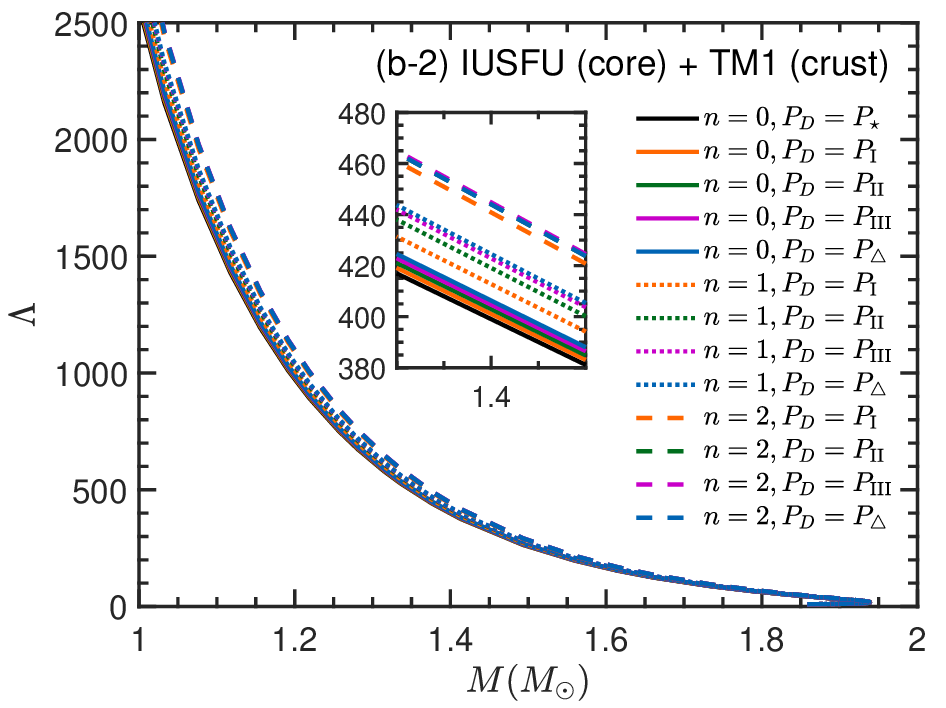}
\end{minipage}
\begin{minipage}{0.325\textwidth}
\includegraphics[scale=0.37]{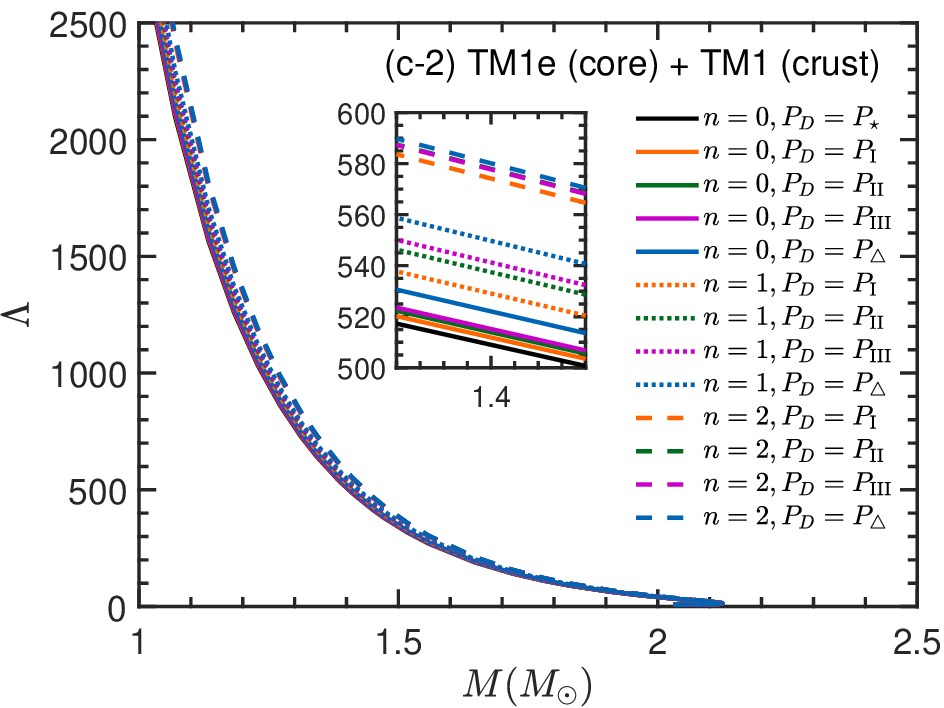}
\end{minipage}
\centering
\begin{minipage}{0.325\textwidth}
\includegraphics[scale=0.37]{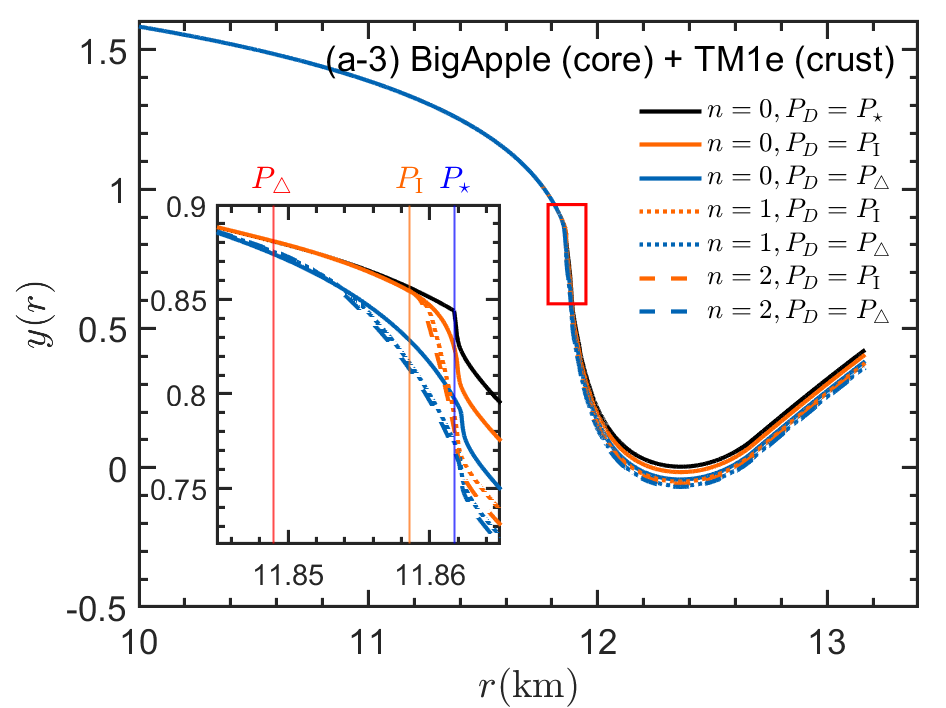}
\end{minipage}
\begin{minipage}{0.325\textwidth}
\includegraphics[scale=0.37]{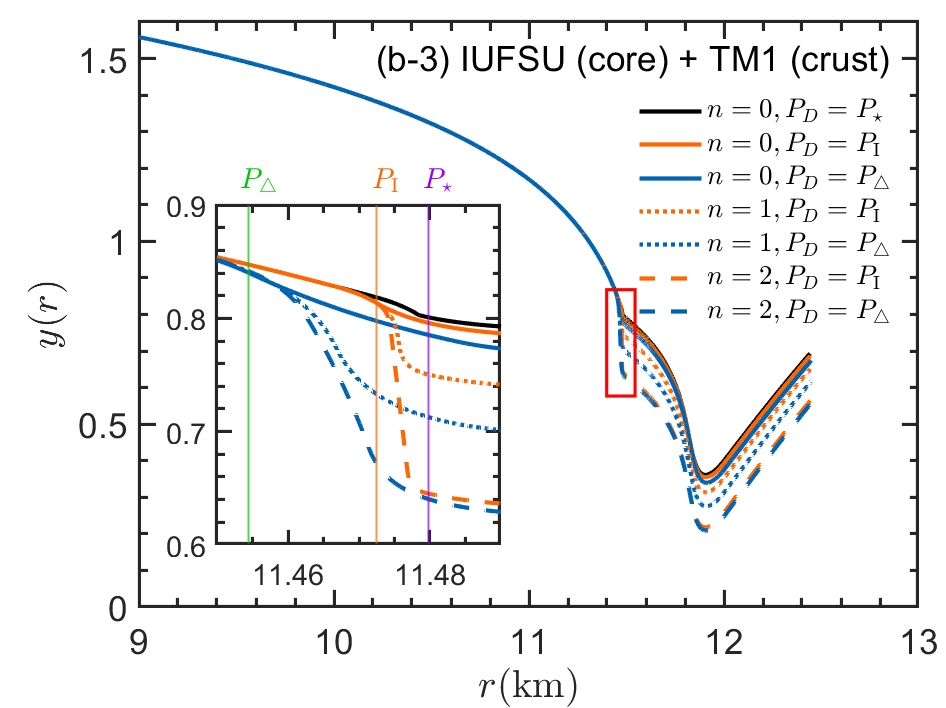}
\end{minipage}
\begin{minipage}{0.325\textwidth}
\includegraphics[scale=0.37]{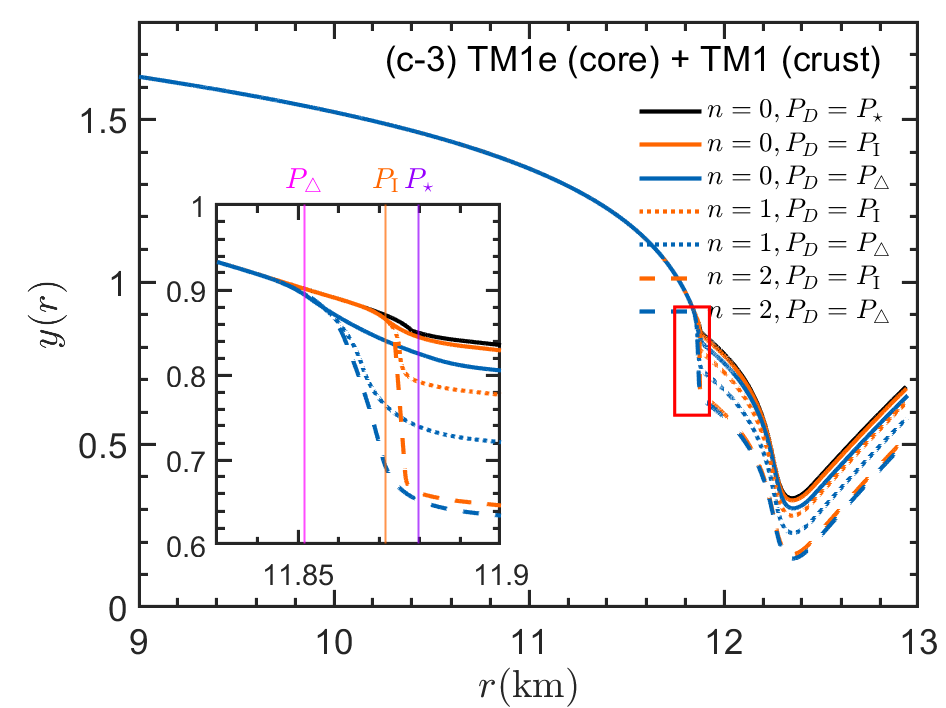}
\end{minipage}
\caption{(a,b,c-1) Love number $k_2$ and (a,b,c-2) tidal deformability $\Lambda$ 
as a function of the neutron-star mass $M$ predicted by the direct-connection EOSs shown in Fig.~\ref{eosdc}. 
The insets show more details for canonical neutron stars with masses around 1.4$\Msun$. 
(a,b,c-3) The $y(r)$ profiles as given in Eq.~\eqref{eq:yr} for a 1.4$\Msun$ neutron star. 
The insets show the results in the direct connection region. 
The pressures corresponding to the vertical lines are consistent with those in Fig.~\ref{eosdc}. }
\label{dctidal}
\end{figure*}
\begin{figure*}[tbp]
\centering
\begin{minipage}{0.325\textwidth}
\includegraphics[scale=0.372]{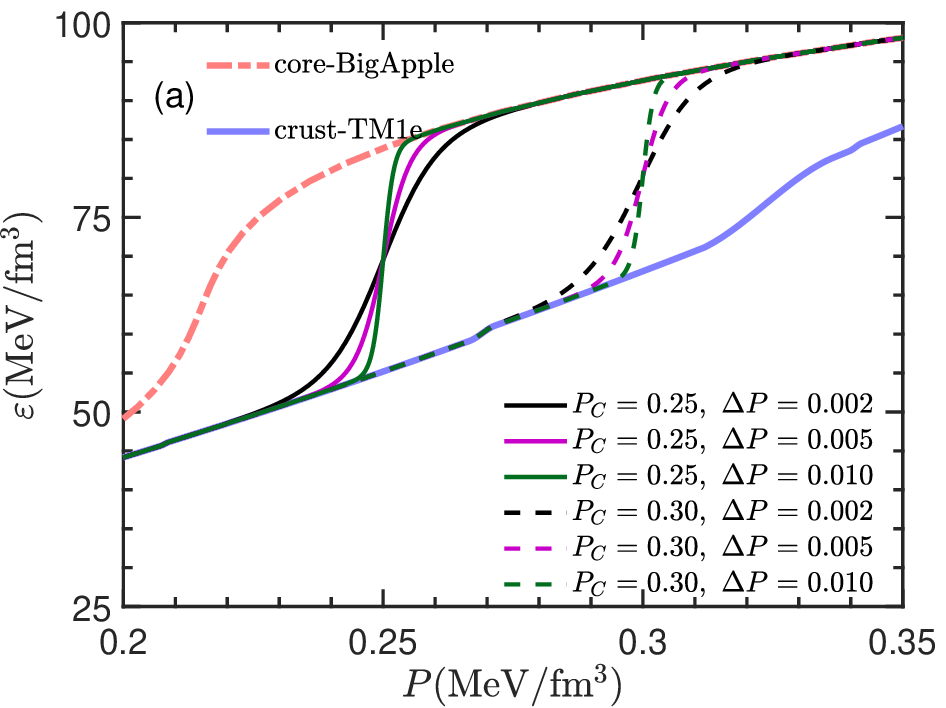}
\end{minipage}
\begin{minipage}{0.325\textwidth}
\includegraphics[scale=0.372]{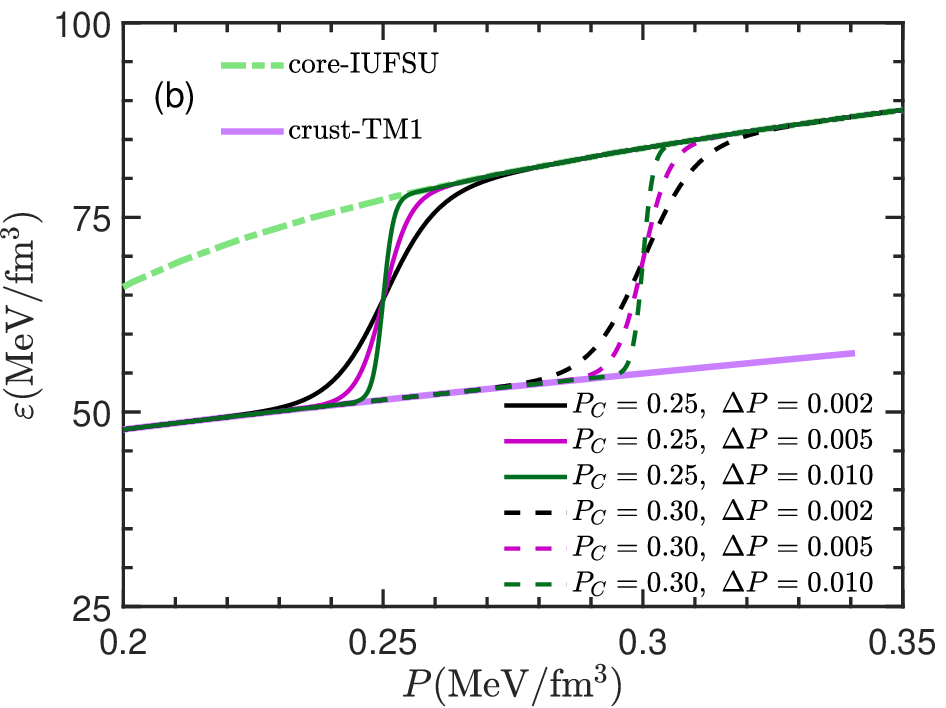}
\end{minipage}
\begin{minipage}{0.325\textwidth}
\includegraphics[scale=0.372]{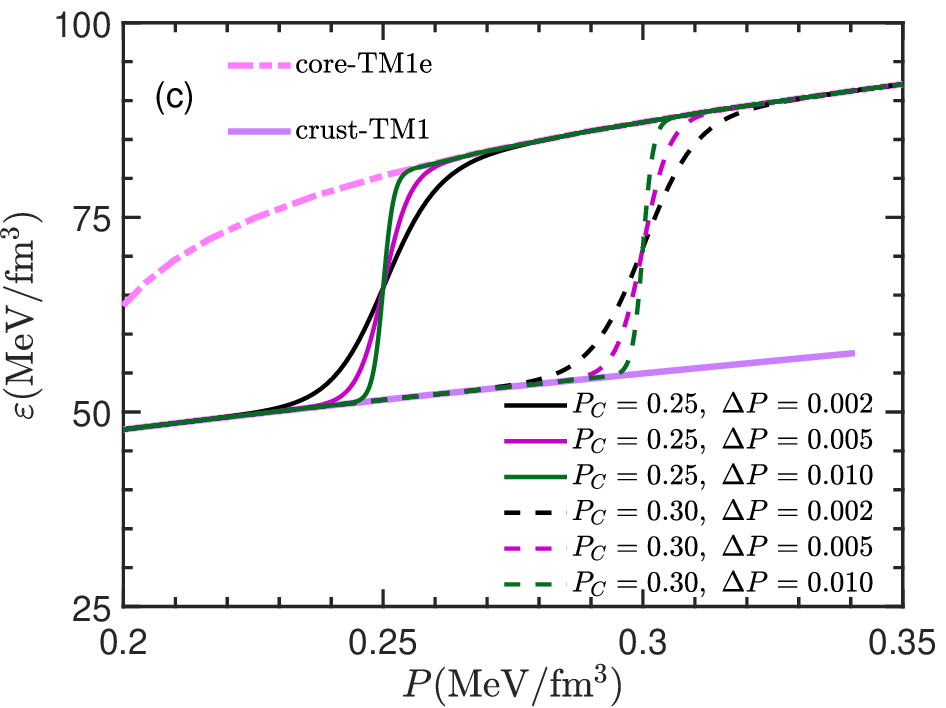}
\end{minipage}
\caption{Profiles of the crossover-connection EOSs. 
(a) shows the results for the combination of BigApple (core) + TM1e (crust). 
(b) is for IUFSU (core) + TM1 (crust). 
(c) is for TM1e (core) + TM1 (crust).}
\label{eoscr}
\end{figure*}
Based on these connection methods, we construct three sets of EOS 
for each combination of crust and core EOSs, as shown in the top ($n=0$), 
middle ($n=1$), and bottom ($n=2$) panels of Fig.~\ref{eosdc}. 
It is interesting to examine how sensitive the properties of neutron stars, 
such as the mass-radius relation and tidal deformability, are to
the connection procedure.

In Fig.~\ref{mrdc}, we show the mass-radius relations obtained using these EOSs.
It is seen that, for each combination, the mass-radius relations 
obtained with different $P_D$ and $n$ are almost indistinguishable.
There are very tiny variations for the radii of canonical 1.4$\Msun$
neutron stars, as shown in the insets. 
For the combination (a) BigApple (core) + TM1e (crust), 
the maximum difference in the radius of a 1.4$\Msun$ neutron star ($R_{1.4}$) 
is less than 10$^{-3}$ km. For the other two combinations, the maximum difference 
in $R_{1.4}$ does not exceed 10$^{-2}$ km.
This implies that the differences in the crust-core connections, as illustrated 
in Fig.~\ref{eosdc}, have a negligible impact on the mass-radius relations of neutron stars. 
However, the influence of the crust-core connections on the tidal deformability of neutron stars 
may be noticeable.

In Fig.~\ref{dctidal}(a-1), (b-1), (c-1), the tidal Love number $k_2$ is shown 
as a function of the neutron-star mass $M$, using the EOSs presented in Fig.~\ref{eosdc}. 
It is seen that the values 
of $k_2$ increase with $M$, reach a maximum value around 0.8$-$1.0$\Msun$, and then 
decrease as $M$ continues to increase until reaching the maximum mass. 
The details shown in the insets indicate that variations in the crust-core connections 
do affect the values of $k_2$ for canonical neutron stars with masses around 1.4$\Msun$. 
It is found that the values of $k_2$ increase with increasing $n$ and $P_D$. 
Similar behaviors are also observed in the tidal deformability, 
$\Lambda$, plotted in Fig.~\ref{dctidal}(a-2), (b-2), (c-2). 
As shown in the insets, the variations in the crust-core connections lead to noticeable 
differences in the tidal deformability of a 1.4$\Msun$ neutron star ($\Lambda_{1.4}$). 
The maximum difference in $\Lambda_{1.4}$ is approximately 35 
for the combination of BigApple (core) + TM1e (crust), and about 50 and 80 
for IUFSU (core) + TM1 (crust) and TM1e (core) + TM1 (crust), respectively.
These results show that when no additional data points are introduced into the connection region, 
the tidal deformability is sensitive to the choice of $P_D$. By adding extra data points to 
the connection region, the sensitivity to $P_D$ can be reduced, while these differences are still visible
(see Table~\ref{ti14},~\ref{IUFSU-1.4},~\ref{IUFSU-1.4}). 
Because the variations in $P_D$ and $n$ do not cause significant changes 
in the mass-radius relation, the compactness $C$ is consequently insensitive to these connections. 
Therefore, the differences in $\Lambda_{1.4}$ are mainly attributed to variations in $k_2$, 
as described in Eq.~(\ref{eq:lambda}).
According to Eq.~\eqref{eq:k2}, the value of $k_2$ is determined by the compactness parameter $C$ 
and the value of $y_R$. Consequently, the differences in $k_2$ should be attributed to 
varying behaviors of $y_R$.
In Fig.~\ref{dctidal}(a-3), (b-3), (c-3), we plot several profiles 
of $y(r)$ for a 1.4$\Msun$ neutron star.
It is seen that $y(r)$, as a function of the radial coordinate $r$, decreases smoothly in 
the core region and varies rapidly beyond the crust-core interface, which is indicated by 
the red rectangle. The value of $y(r)$ reaches a minimum inside the crust and then increases 
until reaching the neutron-star surface at radius $R$, where $y_R$ is the value of $y(r)$. 
Taking the combination of BigApple (core) + TM1e (crust) as an example, 
all curves in Fig.~\ref{dctidal}(a-3) are almost identical for $r<\ $11.85 km,
because of the same core EOS applied in the calculations at this stage.
However, due to the variations in the crust-core connections, for $r>\ $11.85 km,
differences among the curves can be observed and gradually become more visible, 
eventually leading to significantly different values of $y_R$. 
To investigate the effects of different crust-core connections on $y_R$, 
we show the details near the crust-core interface in the inset of Fig.~\ref{dctidal}(a-3).
The vertical lines in different colors indicate that the pressure at $r=r_D$ satisfies
$P|_{r=r_D}=P_D$. All calculations in this case employ the TM1e crust EOS for $r>r_{\star}$.
When no additional data points are added with $n=0$, we find that as $P_D$ increases from 
$P_{\star}$ to $P_{\triangle}$, the values of $y(r)$ in the crust-core 
connection region are in decreasing order. Eventually, different values of $y_R$ are obtained. 
This implies that the differences in $y_R$ are mainly attributed to different $P_D$ used.
When additional data points are added in the cases of $n=1$ and $n=2$, the values of $y(r)$ 
in the crust-core connection region are significantly reduced. 
Comparison of the results for $n=0$ to those for $n=1$ and $n=2$ shows that the differences 
in $y(r)$ curves are more sensitive to the choice of $P_D$ in the set of $n=0$. 
Increasing the number of additional data points can effectively reduce this sensitivity. 
Similar behaviors are also observed in the other two combinations of 
the crust and core EOSs.

The effects of changes in $n$ and $P_D$ on the tidal deformability are mediated through 
their impact on $y(r)$. To be more specific, the impact on $y(r)$ is caused by changes 
in the sound speed squared $c^2_s=\partial P/\partial\varepsilon$ in the crust-core 
connection region, 
which can be seen in Eq.~\eqref{eq:qr}. 
The partial derivative term $\partial P/\partial\varepsilon$ in $Q(r)$ appears 
in the denominator and must 
satisfy $\partial P/\partial\varepsilon\le\ $1 due to causality constraints. 
Consequently, even slight variations in this term can lead to significant changes 
in $y(r)$ as shown in the insets of Fig.~\ref{dctidal}, thereby influencing the 
final value of $y_R$. By increasing the number of additional data points, we can 
reduce the magnitude of changes in $\partial P/\partial\varepsilon$, 
which in turn decreases the sensitivity to the choice $P_D$. 


\begin{figure*}[tbp]
\centering
\begin{minipage}{0.325\textwidth}
\includegraphics[scale=0.375]{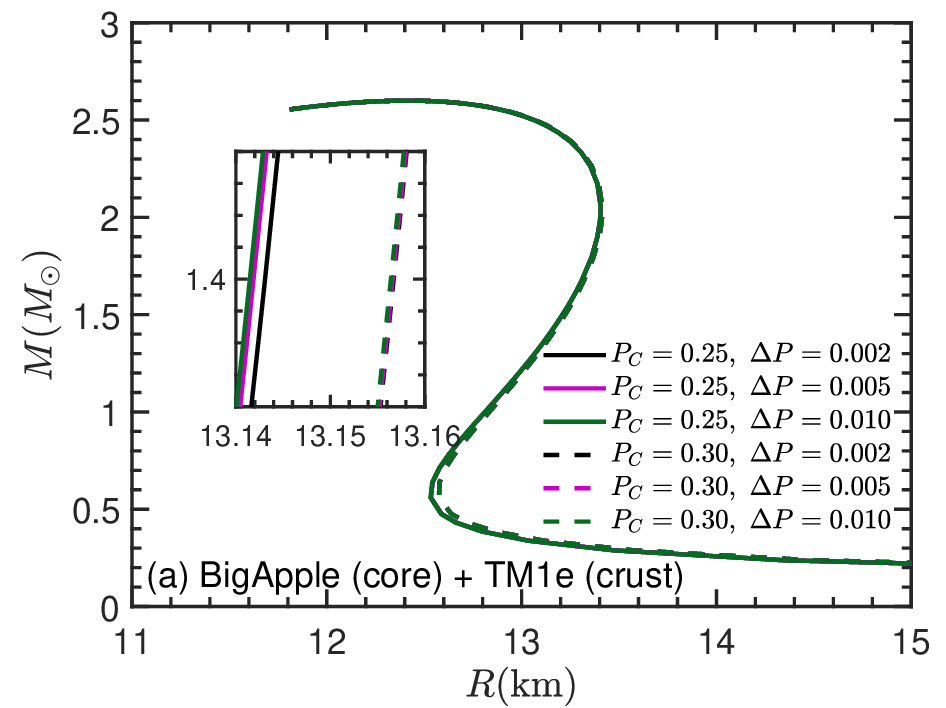}
\end{minipage}
\begin{minipage}{0.325\textwidth}
\includegraphics[scale=0.375]{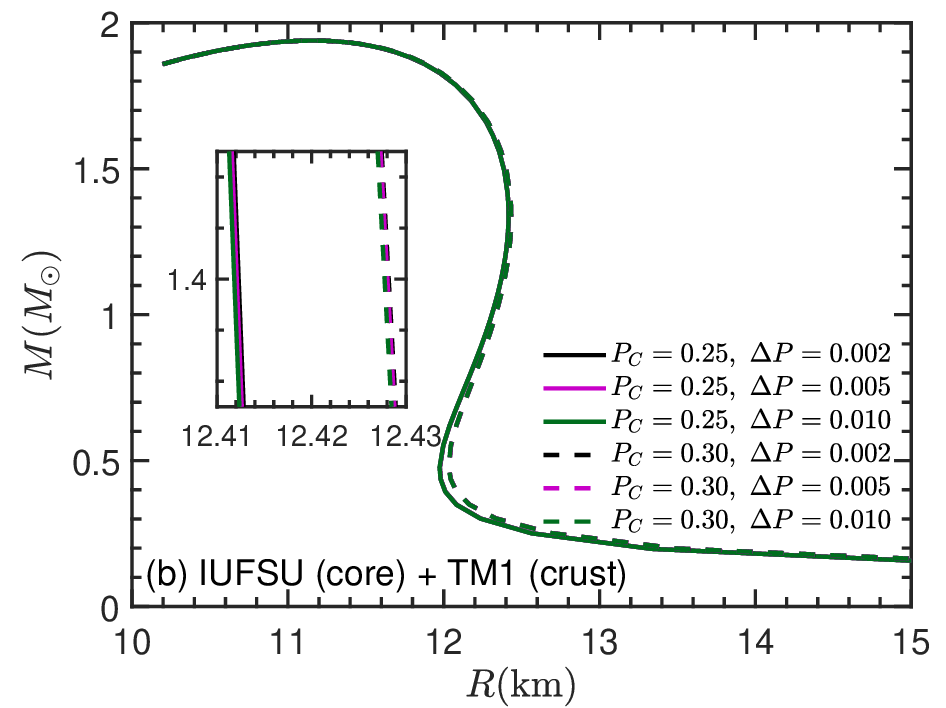}
\end{minipage}
\begin{minipage}{0.325\textwidth}
\includegraphics[scale=0.375]{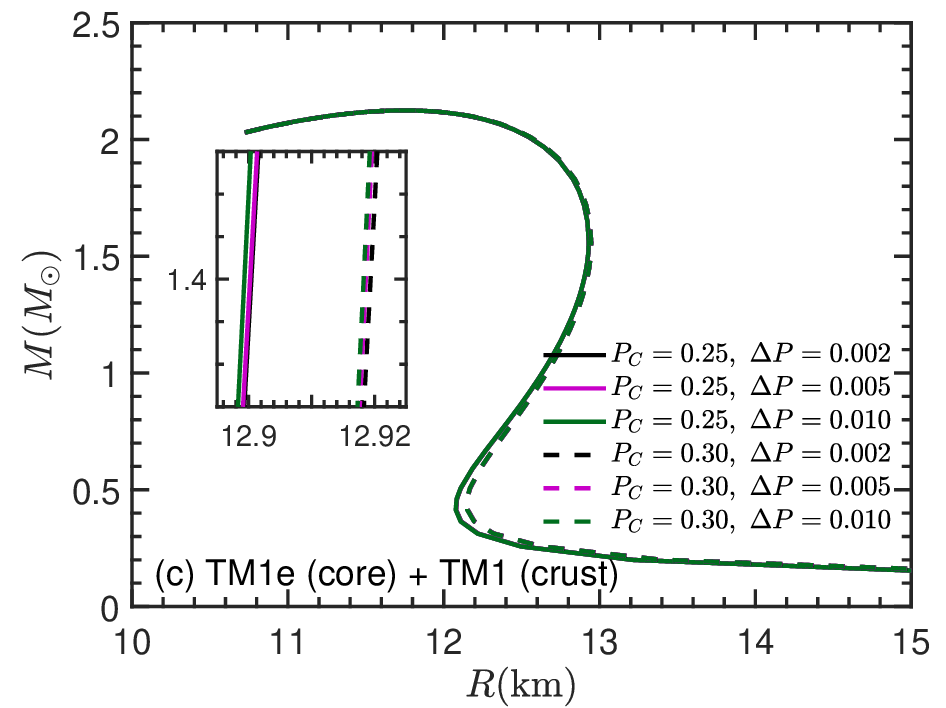}
\end{minipage}
\caption{Mass-radius relations predicted by the crossover-connection EOSs shown in Fig.~\ref{eoscr}. The insets show more details for canonical neutron stars with masses around 1.4$\Msun$.}
\label{mrcr}
\end{figure*}
\begin{figure*}[tbp]
\centering
\begin{minipage}{0.325\textwidth}
\includegraphics[scale=0.37]{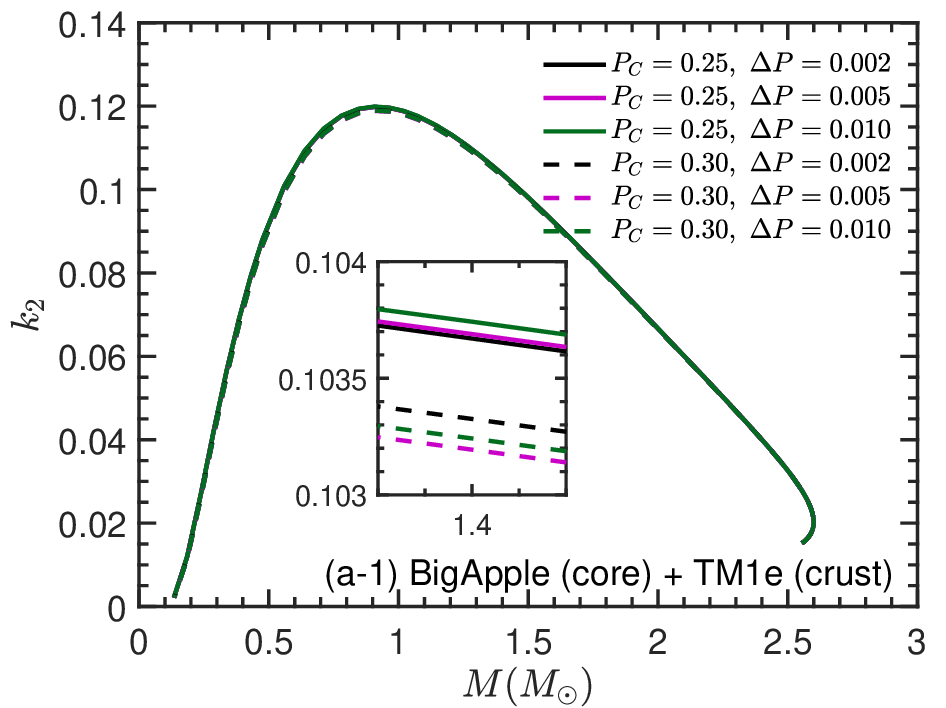}
\end{minipage}
\begin{minipage}{0.325\textwidth}
\includegraphics[scale=0.37]{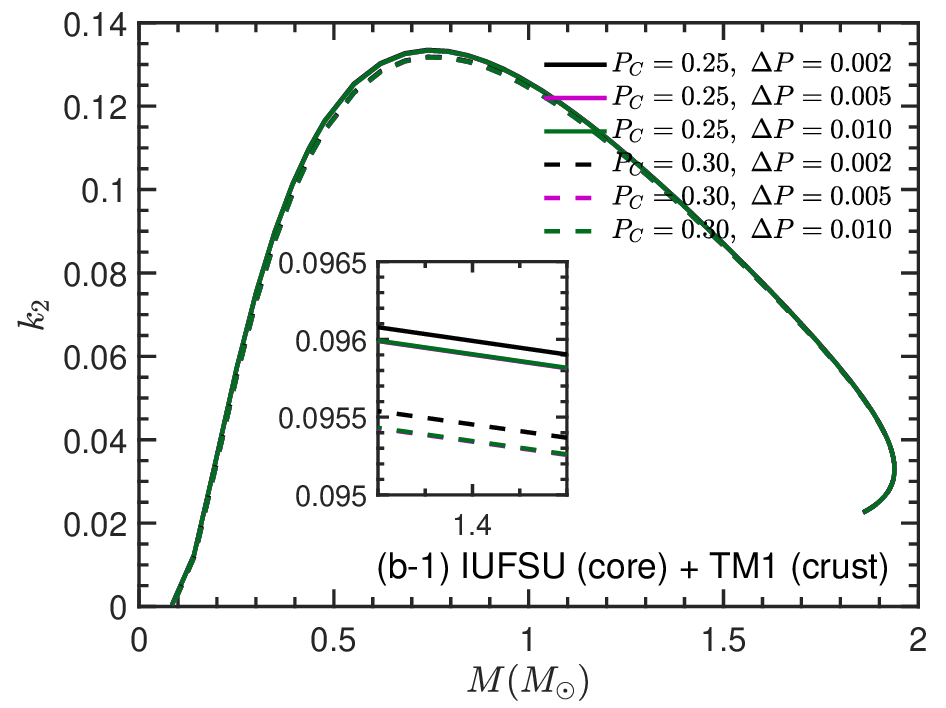}
\end{minipage}
\begin{minipage}{0.325\textwidth}
\includegraphics[scale=0.37]{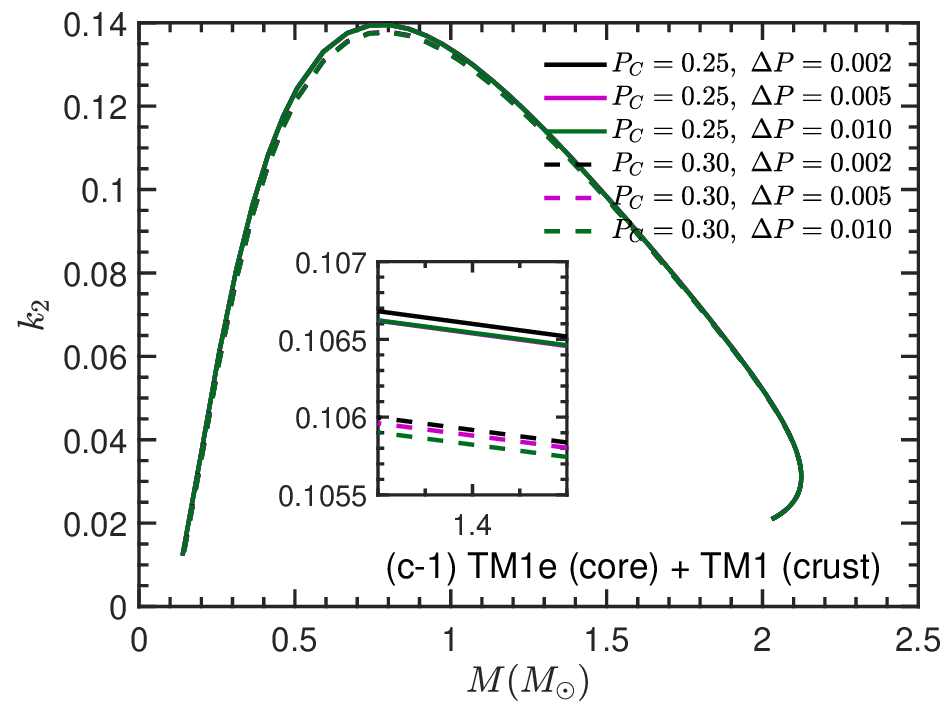}
\end{minipage}
\centering
\begin{minipage}{0.325\textwidth}
\includegraphics[scale=0.37]{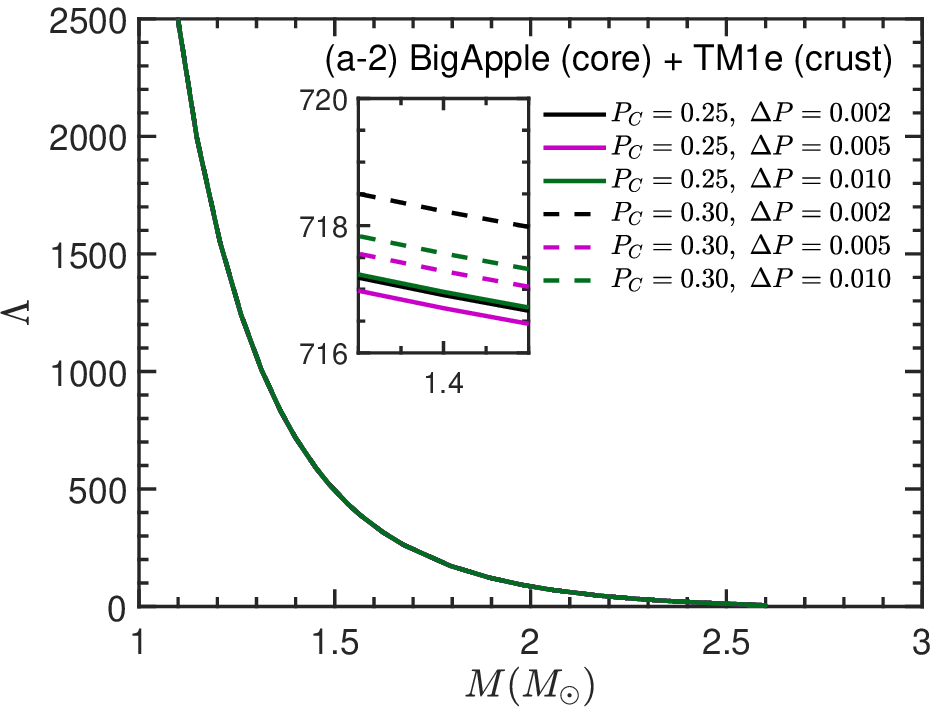}
\end{minipage}
\begin{minipage}{0.325\textwidth}
\includegraphics[scale=0.37]{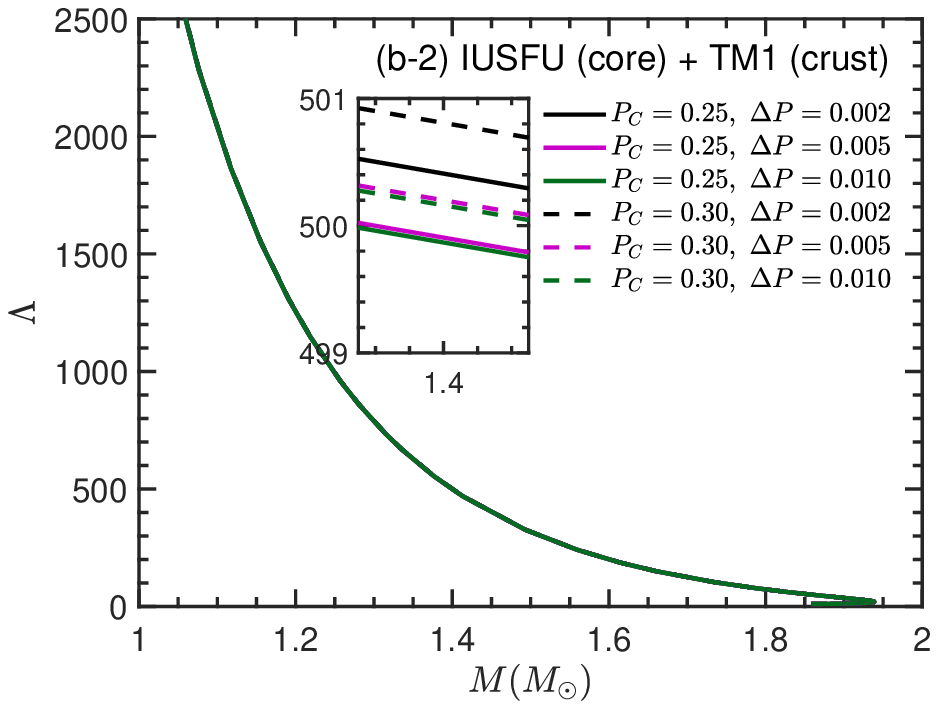}
\end{minipage}
\begin{minipage}{0.325\textwidth}
\includegraphics[scale=0.37]{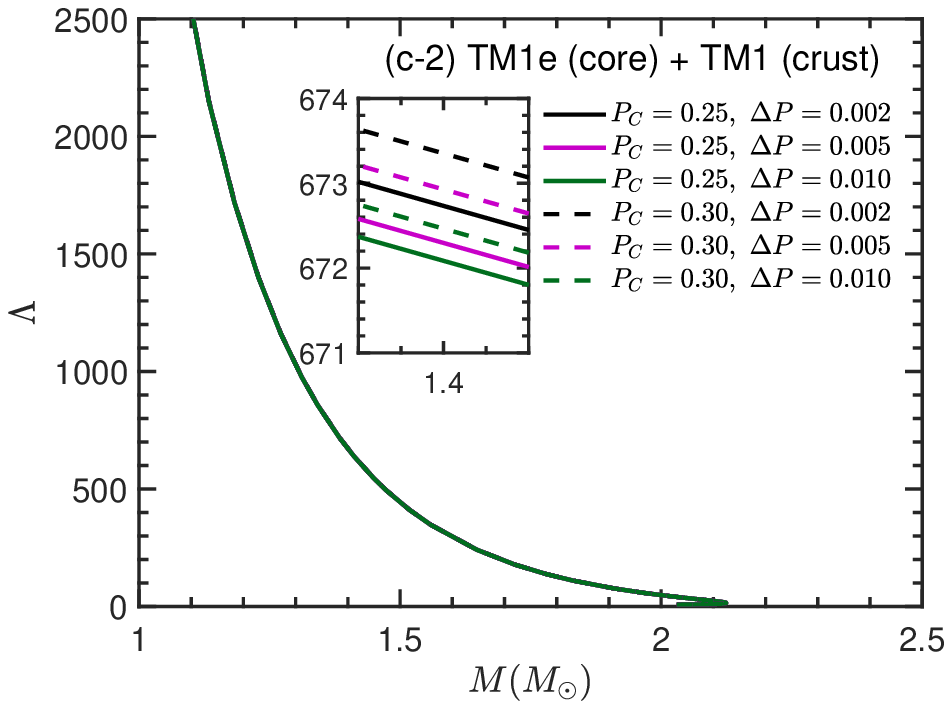}
\end{minipage}
\centering
\begin{minipage}{0.325\textwidth}
\includegraphics[scale=0.37]{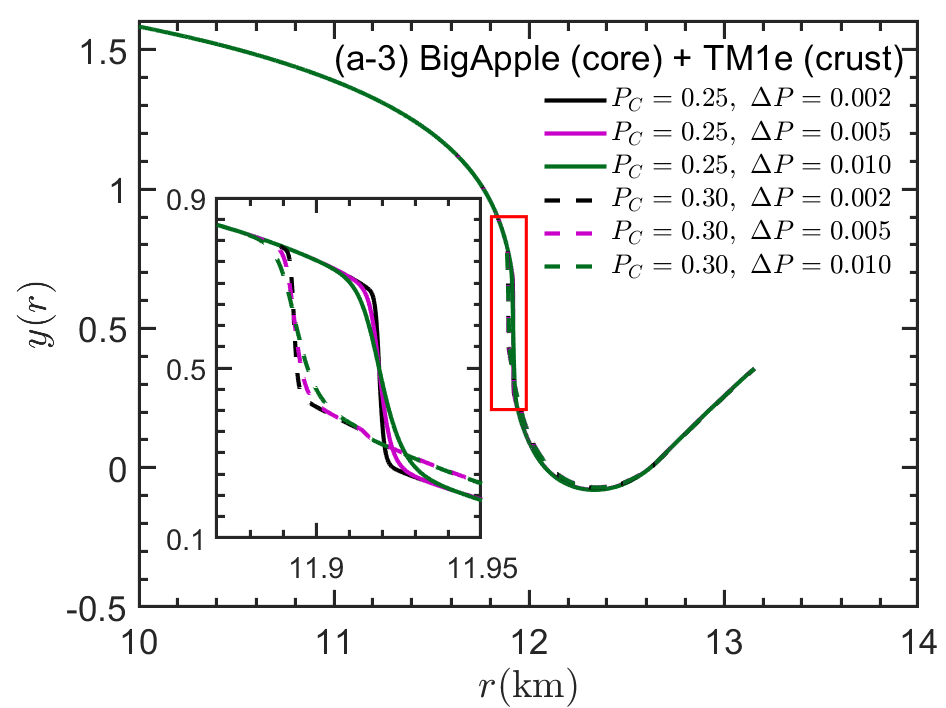}
\end{minipage}
\begin{minipage}{0.325\textwidth}
\includegraphics[scale=0.37]{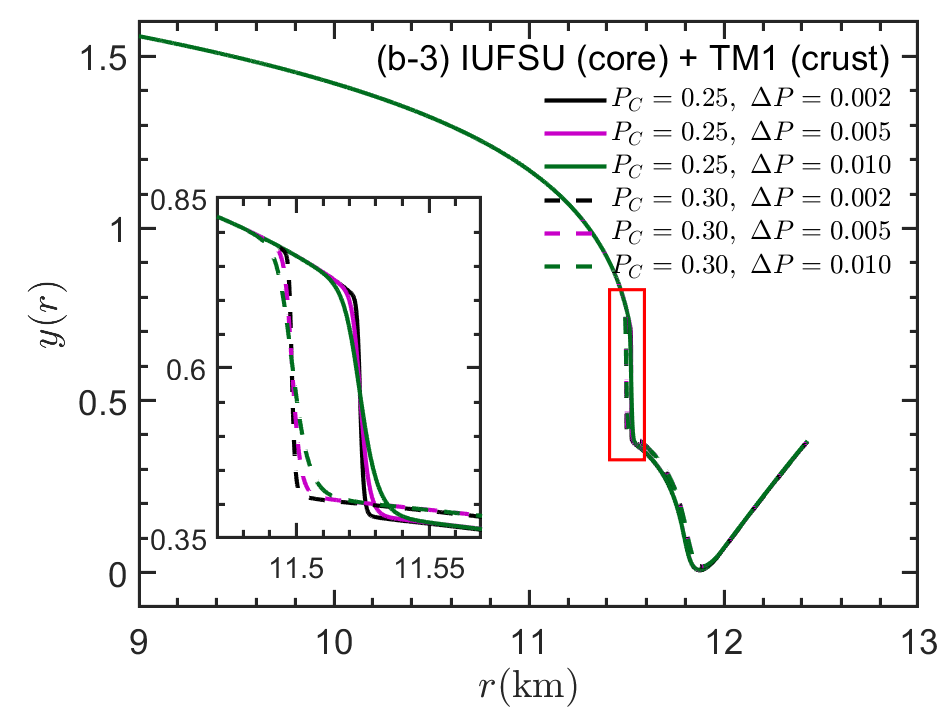}
\end{minipage}
\begin{minipage}{0.325\textwidth}
\includegraphics[scale=0.37]{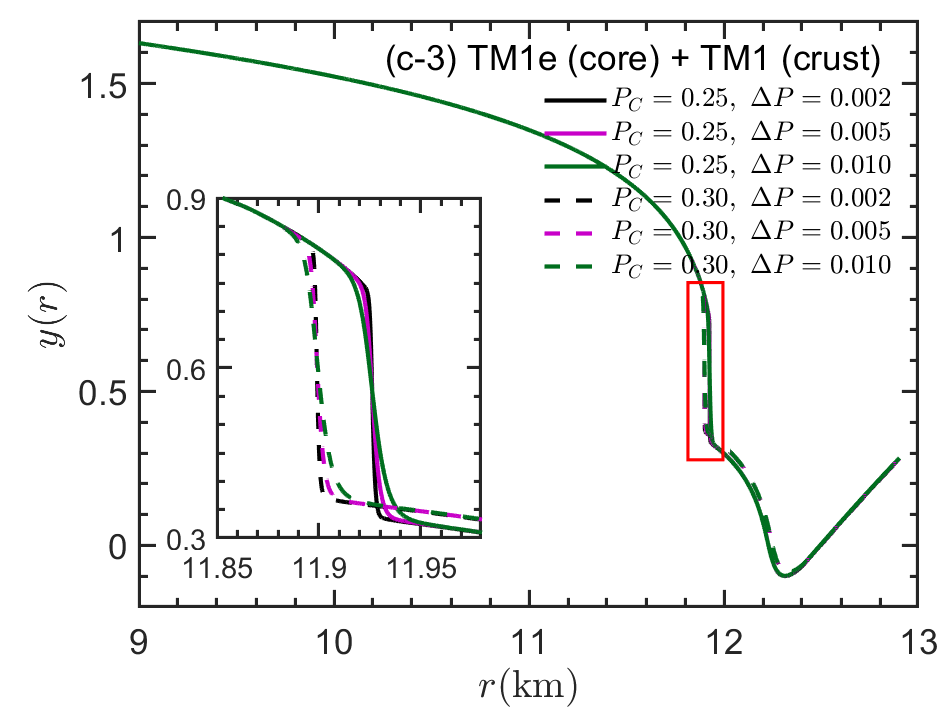}
\end{minipage}
\caption{(a,b,c-1) Love number $k_2$ and (a,b,c-2) tidal deformability $\Lambda$ as 
a function of the neutron-star mass $M$ predicted by the crossover-connection EOSs shown in Fig.~\ref{eoscr}. 
The insets show more details for canonical neutron stars with masses around 1.4$\Msun$. 
(a,b,c-3) The $y(r)$ profiles as given in Eq.~\eqref{eq:yr} for a 1.4$\Msun$ neutron star. 
The insets show the results in the crossover connection region.}
\label{crtidal}
\end{figure*}
\begin{figure*}[tbp]
\centering
\begin{minipage}{0.325\textwidth}
\includegraphics[scale=0.375]{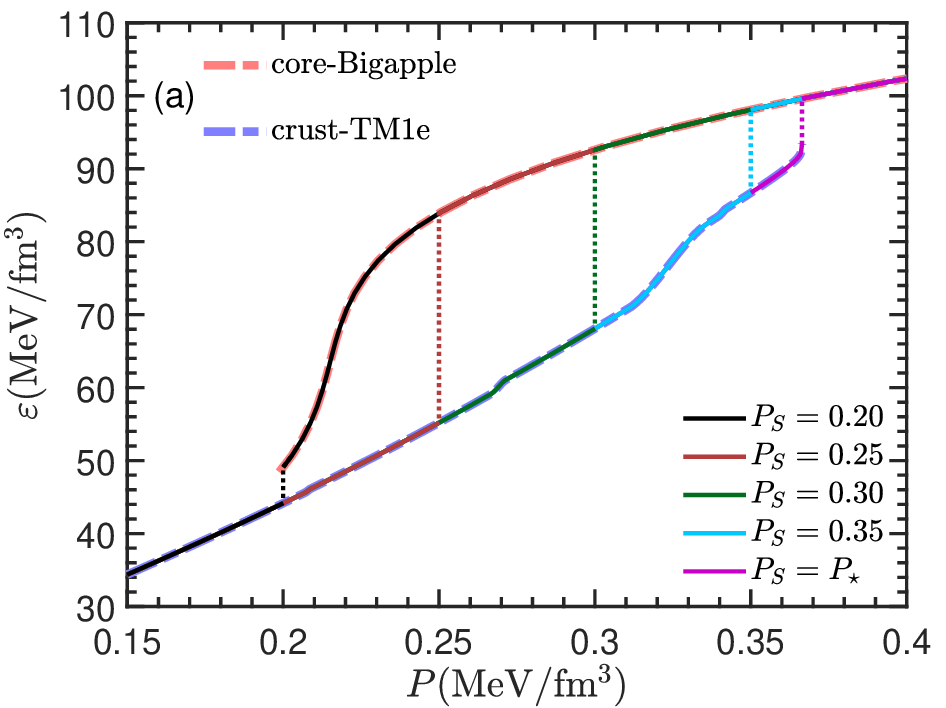}
\end{minipage}
\begin{minipage}{0.325\textwidth}
\includegraphics[scale=0.375]{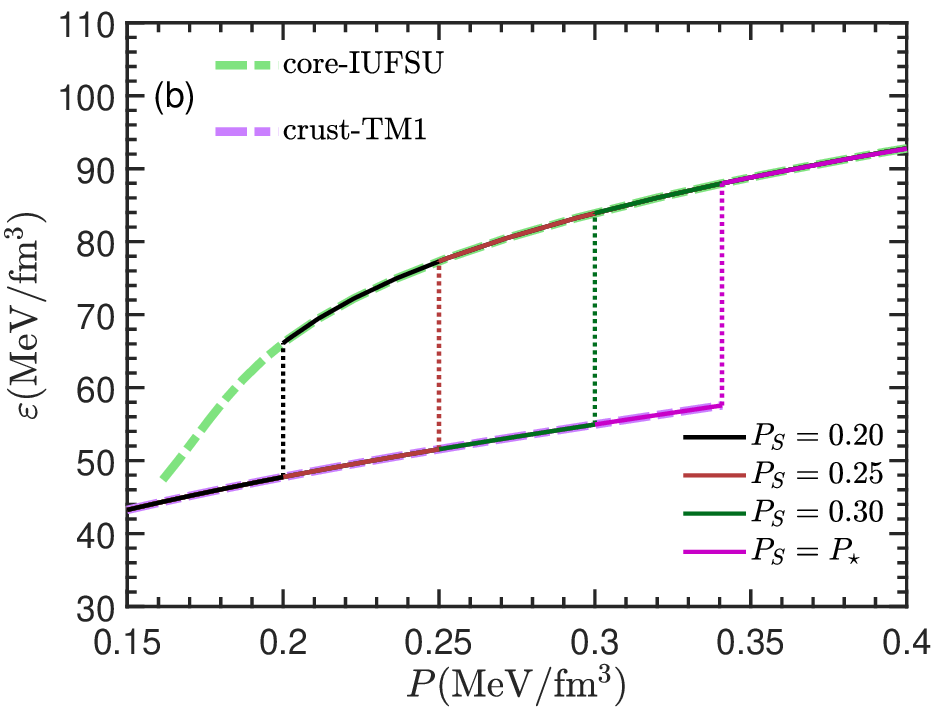}
\end{minipage}
\begin{minipage}{0.325\textwidth}
\includegraphics[scale=0.375]{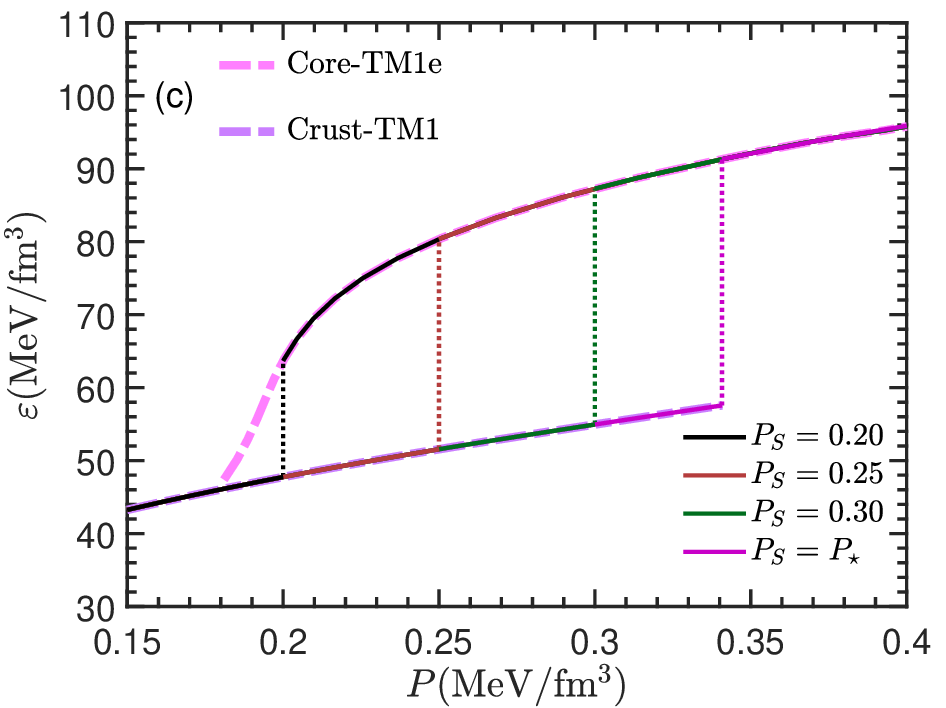}
\end{minipage}
\caption{The EOS profiles in the segmented method. The crust-core transition 
is assumed to occur at $P_S$, which is set to different values. The case of $P_S=P_{\star}$ 
corresponds to the direct connection case with $n=0$ and $P_D=P_{\star}$, where the inner crust EOS is completely accessed.
(a) shows the results for the BigApple (core) + TM1e (crust).
(b) is for the IUFSU (core) + TM1 (crust).
(c) is for the TM1e (core) + TM1 (crust).}
\label{pweos}
\end{figure*}
\subsection{Crossover connection}
\label{sec-cc}

In this section, we conduct a regularized 
calculation to generate a crossover EOS that bridges the inner crust and core segments.
We employ the regularized form as described in~\cite{alfo17} for the energy density in the 
connection region, which is expressed as
\begin{equation}
\begin{aligned}
\varepsilon(P)=\frac{1}{2}&\left[1+\mathrm{tanh}\left(\frac{P-P_C}{\Delta P}\right)\right]\varepsilon_{\mathrm{core}}(P)\\
+\frac{1}{2}&\left[1-\mathrm{tanh}\left(\frac{P-P_C}{\Delta P}\right)\right]\varepsilon_{\mathrm{crust}}(P).
\end{aligned}
\label{intp}
\end{equation}
Here, $P_C$ indicates the critical pressure at which the crust-core phase transition occurs, 
and $\Delta P$ is a parameter used to control the extent of the connection region.
To ensure a smooth connection of the crossover EOS to the original crust and core segments, 
we employ the regularized form given in Eq.~\eqref{intp} until the relative difference
in the energy density $\varepsilon$ is less than 10$^{-3}$ at both ends of the connection 
region. 

In Fig.~\ref{eoscr}, we present several profiles of the crossover-connection EOSs.
The original crust and core segments are distinguished by thicker lines.
To investigate the effects of $P_C$ and $\Delta P$, we consider two critical pressure 
values ($P_C=\ $0.25 and 0.30 MeV$/$fm$^3$) along with three distinct $\Delta P$ 
values ($\Delta P=\ $0.002, 0.005, and 0.010 MeV$/$fm$^3$). 
The curves with $\Delta P=\ $ 0.002 MeV$/$fm$^3$ as shown by the black lines are 
closer to a first-order phase transition, in which sufficient data points are still included 
to ensure a smooth change in the sound speed squared. 
As $\Delta P$ increases, the crossover EOS between the crust and core segments becomes 
smoother, and the crossover region expands.

In Fig.~\ref{mrcr}, we display the mass-radius relations of neutron stars using the EOSs with the 
crossover connections. The neutron-star radii are shown to be insensitive to the value of $\Delta P$. 
Regarding the influence of $P_C$, it is seen that massive neutron stars exhibit almost no difference 
in their radii, whereas canonical neutron stars with masses around 1.4$\Msun$ show small differences 
in radii, as illustrated in the insets of Fig.~\ref{mrcr}. 
The radii with $P_C=\ $0.25 MeV$/$fm$^3$ (solid lines) are slightly smaller than those 
with $P_C=\ $0.30 MeV$/$fm$^3$ (dashed lines). 
In the three combinations of the crust and core EOSs shown in Fig.~\ref{mrcr}, 
the influences of $\Delta P$ and $P_C$ on the mass-radius relations follow a similar trend.

In Fig.~\ref{crtidal}, we show the tidal Love number $k_2$ (top panels) and the dimensionless 
tidal deformability $\Lambda$ (middle panels) as a function of the neutron-star mass $M$. 
These results are obtained using the crossover-connection EOSs shown in Fig.~\ref{eoscr}. 
For the sets with the same $P_C$, the $k_2-M$ curves do not exhibit significant 
divergence. However, between the sets with $P_C=\ $0.25 MeV$/$fm$^3$ (solid lines) 
and $P_C=\ $0.30 MeV$/$fm$^3$ (dashed lines), small differences can be observed, 
as illustrated in the insets of 
Fig.~\ref{crtidal}(a-1), (b-1), (c-1). In each combination of the crust and core EOSs, 
all curves of tidal deformability $\Lambda$ almost completely overlap. 
Furthermore, the largest difference in $\Lambda_{1.4}$ within one EOS combination is found 
to be less than 2, as shown in the insets. 
In Fig.~\ref{crtidal}(a-3), (b-3), (c-3), we show the $y(r)$ profiles for a 1.4$\Msun$ neutron star 
obtained using the crossover-connection EOSs. 
The $y(r)$ curves for the three EOS combinations exhibit a similar trend.
All curves practically coincide in the core region. 
Small differences appear only within the crust-core connection region, as shown in the insets.
The influence of $\Delta P$ is almost negligible, whereas the critical pressure $P_C$ yields 
a visible distinction on the $y(r)$ curves. 
However, when the $y(r)$ curves extend beyond the crossover region, 
the curves with different $P_C$ get closer, and ultimately result in almost identical values of $y_R$.
We find that the differences in the tidal deformability of a 1.4$\Msun$ neutron star ($\Lambda_{1.4}$) 
obtained using the crossover-connection EOSs are very small, which are much smaller than 
those obtained using the direct-connection EOSs
(see Table~\ref{ti14},~\ref{IUFSU-1.4},~\ref{TM1e-1.4}).
This implies that the crossover connection provides more stable results of $\Lambda_{1.4}$ than the direct connection.

\subsection{Segmented method}
\label{sec-sm}

In the direct and crossover connection procedures described above, we first construct an entire 
EOS data table, and then continuously solve the TOV equation from the center to the surface of 
a neutron star. However, when the discontinuity at the crust-core interface is like a first-order 
phase transition, the sound speed squared $c^2_s=\partial P/\partial\varepsilon$ drops close to zero,
due to the same pressure but different energy densities between the two phases.
The presence of $c^2_s \simeq 0$ in the denominator of Eq.~\eqref{eq:qr} may lead to difficulty 
in solving the equation and induce uncertainties in the prediction of the tidal deformability.
In order to overcome these difficulties, we solve the TOV equation in segments, employing 
appropriate matching conditions at the crust-core interface.
This procedure is referred to as the segmented method in the present work.
We note that appropriate matching conditions should be imposed to ensure continuity and 
consistency between the crust and core regions.

At the crust-core interface, the pressure is continuous,
\begin{equation}
P^{tr}_{\mathrm{core}}=P^{tr}_{\mathrm{crust}}=P_S,
\end{equation}
while the energy density must monotonically decrease,
\begin{equation}
\varepsilon^{tr}_{\mathrm{core}}\ge\varepsilon^{tr}_{\mathrm{crust}}.
\end{equation}
We denote the radius and enclosed mass at the crust-core transition point inside the neutron star by,
\begin{equation}
\begin{aligned}
r^{tr}_{\mathrm{core}}&=r^{tr}_{\mathrm{crust}}=r_S,\\
M^{tr}_{\mathrm{core}}&=M^{tr}_{\mathrm{crust}}=M_S.
\end{aligned}
\label{eq:rs}
\end{equation}
According to Eq.~\eqref{eq:yr} with $c^2_s\simeq 0$, the matching condition of $y(r)$ at the crust-core
interface is expressed as~\cite{post10,han19}
\begin{equation}
y^{tr}_{\mathrm{crust}}=y^{tr}_{\mathrm{core}}-\frac{4\pi r^3_{S}(\varepsilon^{tr}_{\mathrm{core}}-\varepsilon^{tr}_{\mathrm{crust}})}{M_S}.
\label{eq:sy}
\end{equation}
The advantage of the segmented method is that no additional modifications are made to the 
original crust and core EOS data. A critical parameter in this method is the transition 
pressure $P_S$, where the crust-core transition is assumed to occur. We compare results 
for $P_S=\ $0.20, 0.25, 0.30, 0.35 MeV$/$fm$^3$ and $P_{\star}$ in the 
segmented method.
\begin{figure*}[tbp]
\centering
\begin{minipage}{0.325\textwidth}
\includegraphics[scale=0.375]{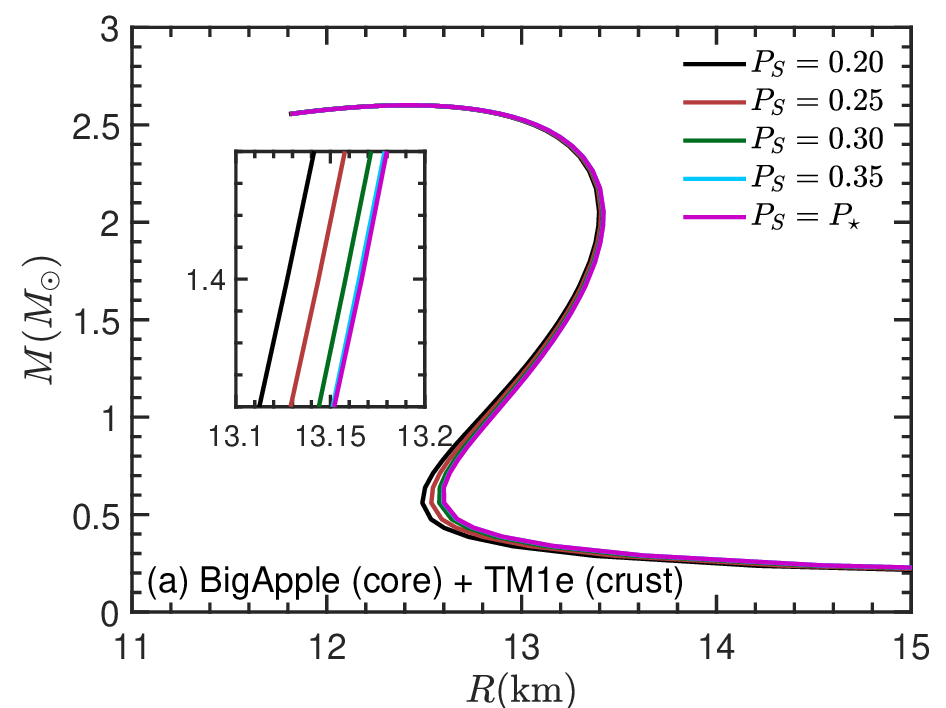}
\end{minipage}
\begin{minipage}{0.325\textwidth}
\includegraphics[scale=0.375]{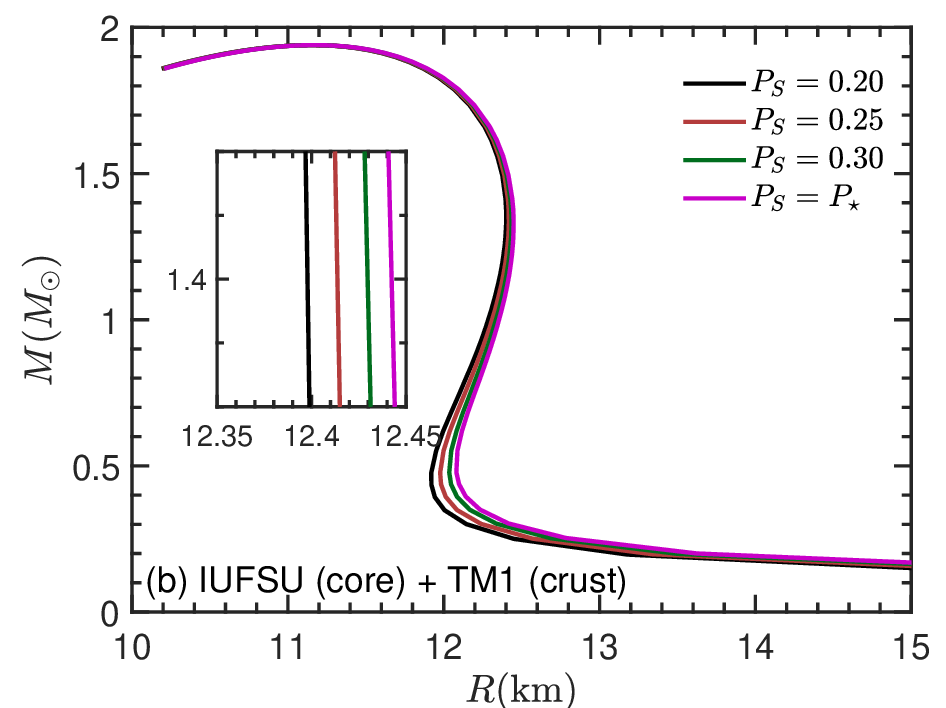}
\end{minipage}
\begin{minipage}{0.325\textwidth}
\includegraphics[scale=0.375]{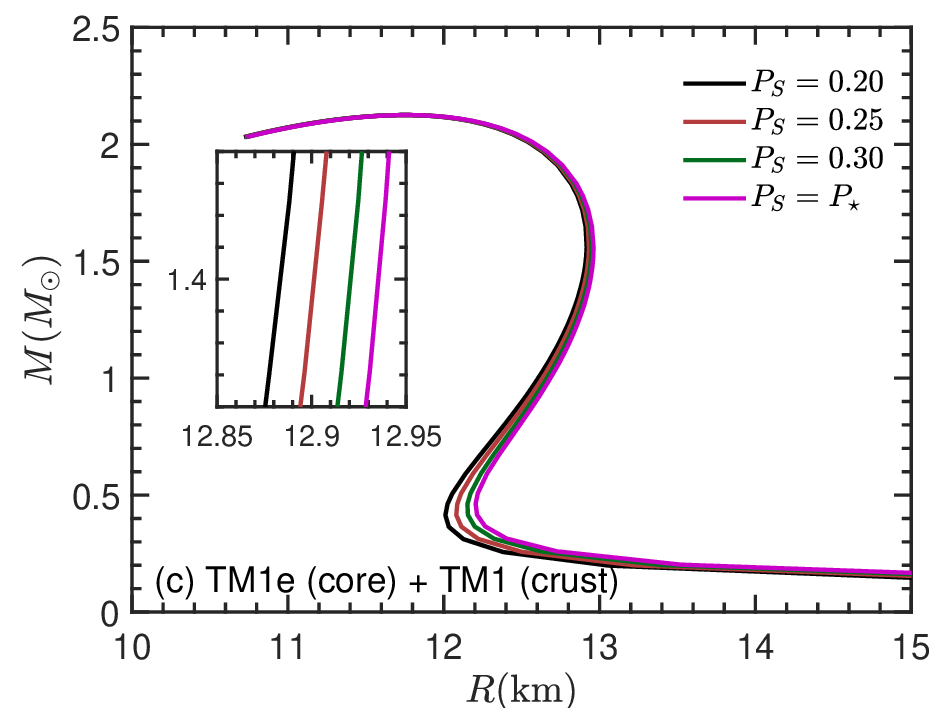}
\end{minipage}
\caption{Mass-radius relations predicted by the segmented method. 
The insets show more details for canonical neutron stars with masses around 1.4$\Msun$.}
\label{pwmrm}
\end{figure*}
\begin{figure*}[tbp]
\centering
\begin{minipage}{0.325\textwidth}
\includegraphics[scale=0.37]{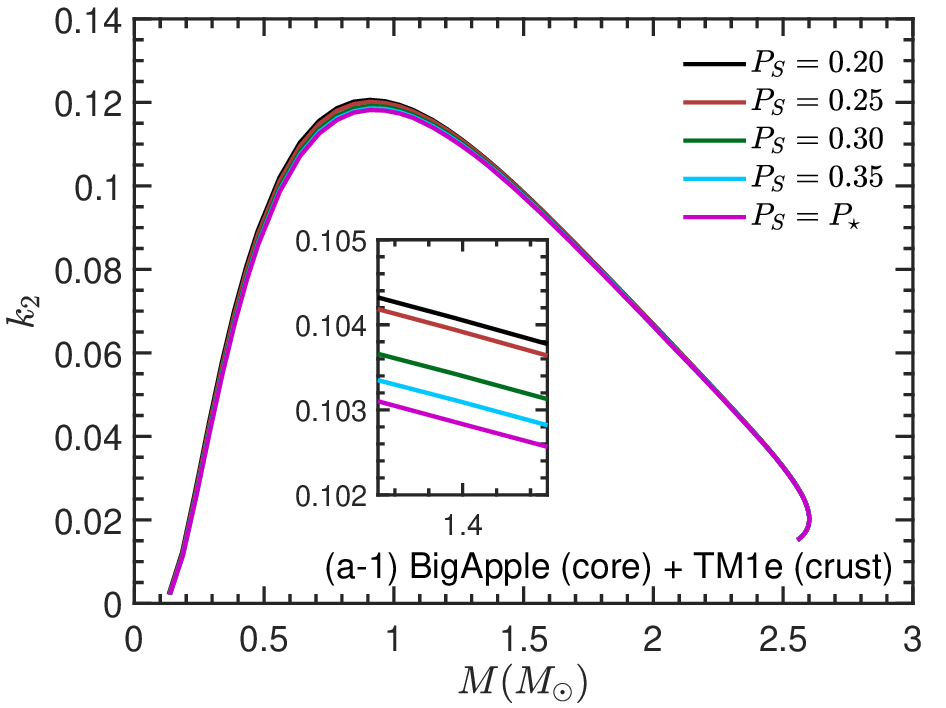}
\end{minipage}
\begin{minipage}{0.325\textwidth}
\includegraphics[scale=0.37]{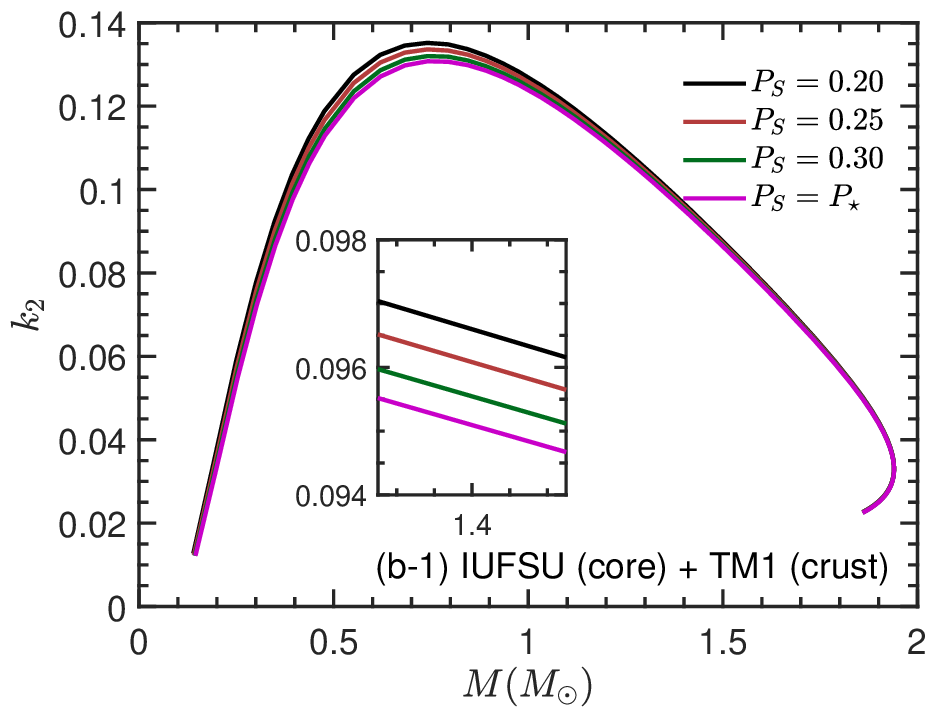}
\end{minipage}
\begin{minipage}{0.325\textwidth}
\includegraphics[scale=0.37]{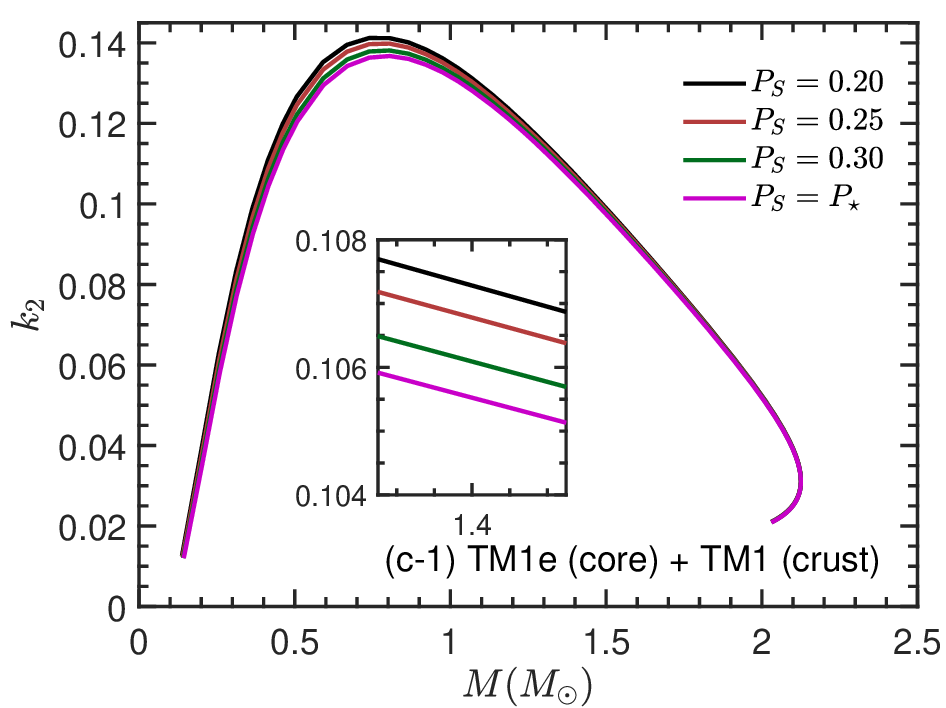}
\end{minipage}
\centering
\begin{minipage}{0.325\textwidth}
\includegraphics[scale=0.37]{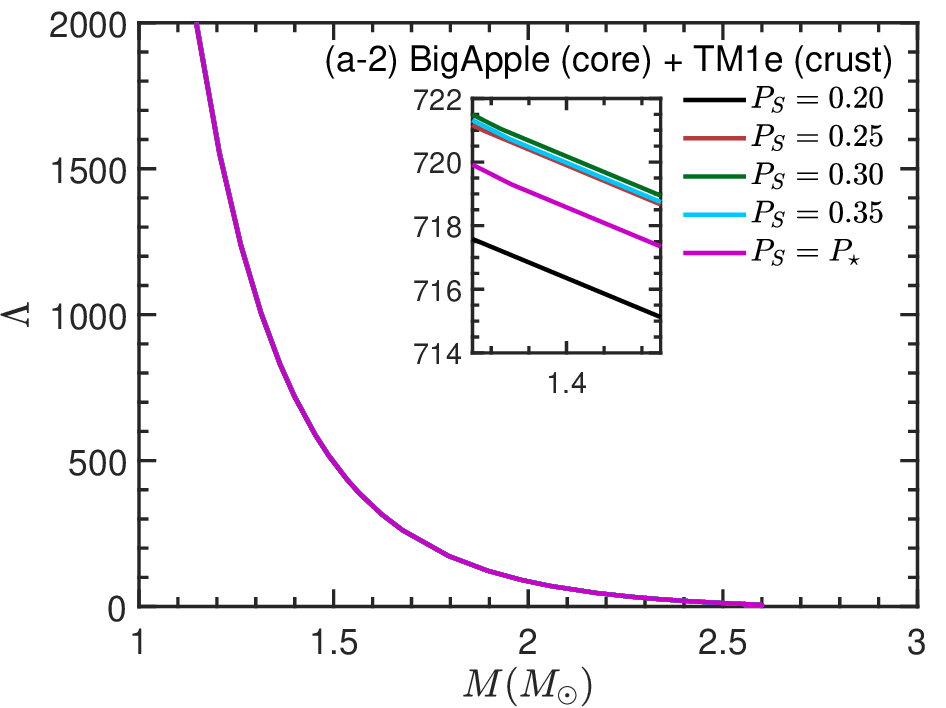}
\end{minipage}
\begin{minipage}{0.325\textwidth}
\includegraphics[scale=0.37]{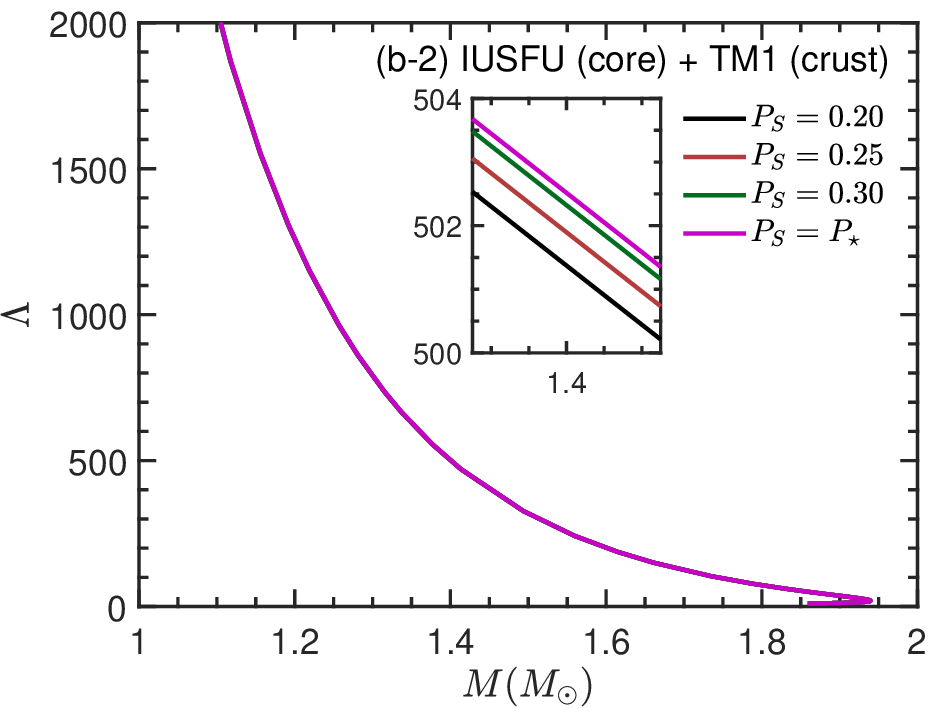}
\end{minipage}
\begin{minipage}{0.325\textwidth}
\includegraphics[scale=0.37]{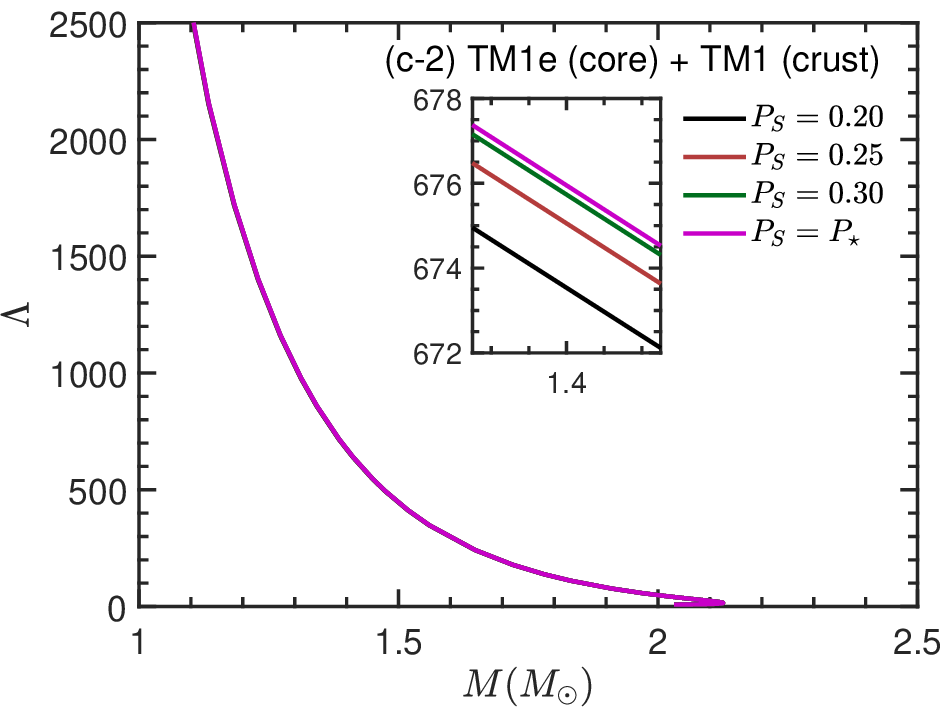}
\end{minipage}
\centering
\begin{minipage}{0.325\textwidth}
\includegraphics[scale=0.37]{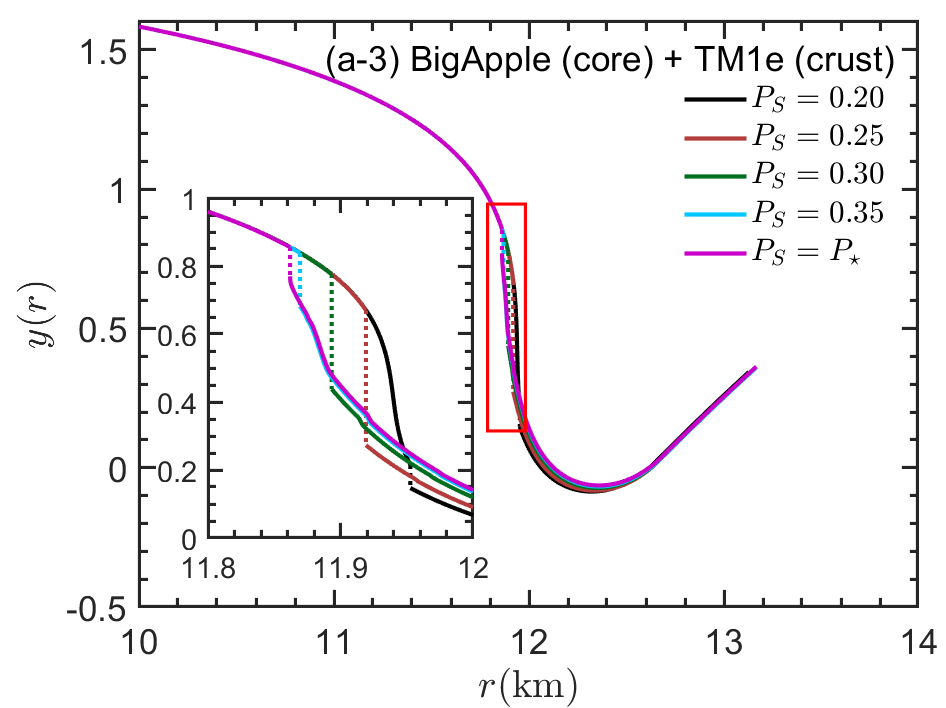}
\end{minipage}
\begin{minipage}{0.325\textwidth}
\includegraphics[scale=0.37]{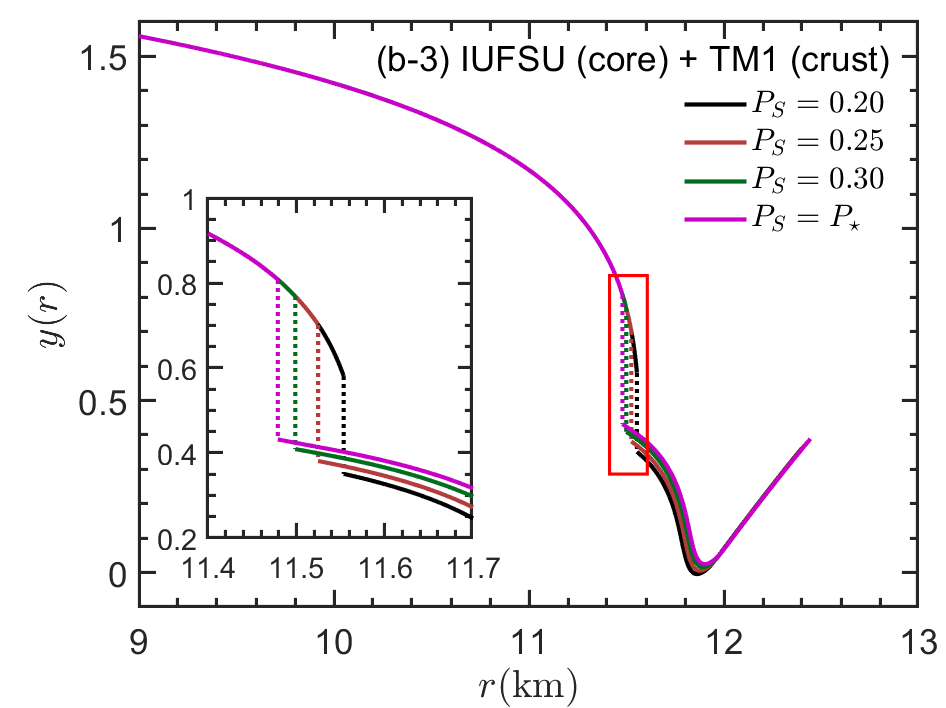}
\end{minipage}
\begin{minipage}{0.325\textwidth}
\includegraphics[scale=0.37]{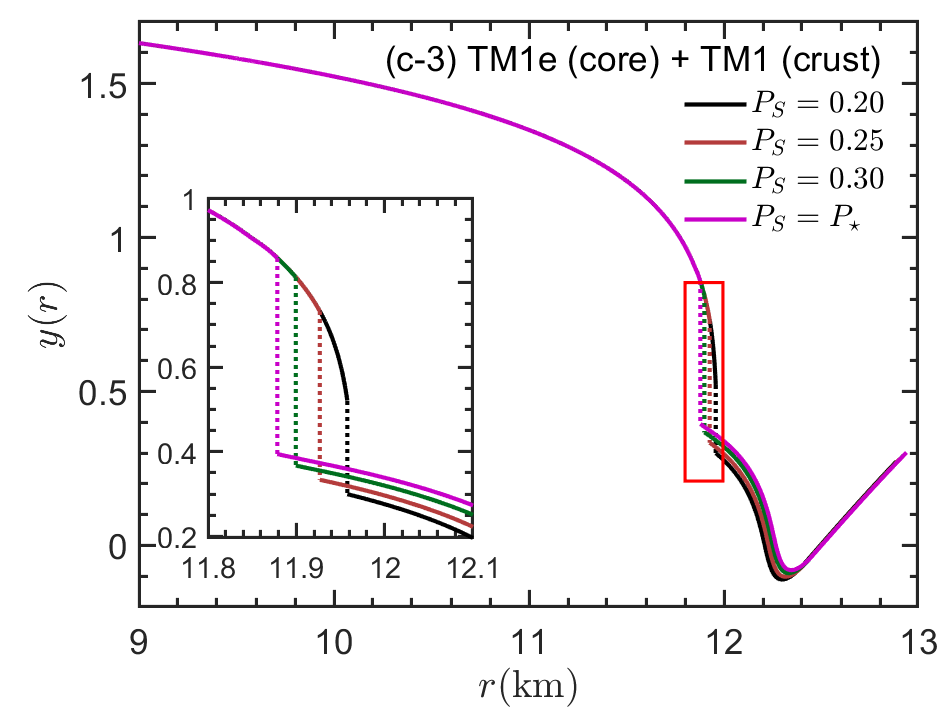}
\end{minipage}
\caption{(a,b,c-1) Love number $k_2$ and (a,b,c-2) tidal deformability $\Lambda$ as 
a function of the neutron-star mass $M$ predicted by the segmented method. 
The insets show more details for canonical neutron stars with masses around 1.4$\Msun$. 
(a,b,c-3) The $y(r)$ profiles as given in Eq.~\eqref{eq:yr} for a 1.4$\Msun$ neutron star. 
The insets show the results in the region where the segmented method is applied. }
\label{pwtidal}
\end{figure*}

In Fig.~\ref{pweos}, we show the EOS profiles in the segmented method with various $P_S$ values. 
The core and crust segments are connected by a vertical dotted line, analogous to the characteristics 
of a first-order phase transition. 
We note that the critical pressures $P_{\star}$ of the TM1 and TM1e models
are about 0.341 and 0.366 MeV$/$fm$^3$, respectively. 
Consequently, the highest $P_S$ shown in Figs.~\ref{pweos}(b) and~\ref{pweos}(c) with the TM1 crust
is $P_{\star}=\ $0.341 MeV$/$fm$^3$, while any $P_S>P_{\star}$ (e.g., $P_S=\ $0.35 MeV$/$fm$^3$)
is excluded.

By solving the TOV equation in segments, we obtain the properties of neutron stars. 
In Fig.~\ref{pwmrm}, we present the mass-radius relations predicted by the segmented method. 
It is noteworthy that changing the value of $P_S$ is equivalent to altering the crust-core transition point. 
This alteration does not significantly affect the radii of neutron stars, especially for massive stars.
There are small differences observed in the radii of canonical neutron stars with masses 
around 1.4$\Msun$, as shown in the insets. It is seen that with increasing $P_S$, 
the radius of a 1.4$\Msun$ neutron star slightly increases, 
while the largest difference in $R_{1.4}$ within one EOS combination
is less than 0.05 km (see Table~\ref{ti14},~\ref{IUFSU-1.4},~\ref{TM1e-1.4}).

\begin{table*}[htbp]
	\begin{center}
		\caption{Properties of a 1.4$\Msun$ neutron star obtained using the direct connection procedure, 
			crossover connection procedure, and segmented method. $\Delta R_{1.4}$, $\Delta k_{2}$, and $\Delta\Lambda_{1.4}$ 
			represent the discrepancies between the direct/crossover connection and the segmented method. 
			The results of the direct connection are compared with those for $P_S=P_{\star}$, 
			and the results of the crossover connection are compared with those for $P_S=P_C$. The results in this table are obtained with 
            the BigApple (core) + TM1e (crust) combination.}
		\begin{tabular}{c|cccccccccccccccccccccccccc}
			\hline\hline
			EOS& \multicolumn{8}{c}{\textbf{BigApple (core) + TM1e (crust)}}\\
			\hline\hline
			\multirow{15}{*}{\makecell[c]{Direct connection}}
			&$P_D$&$n$&  $R_{1.4}$  & $k_2$ & $\Lambda_{1.4}$&$\Delta R_{1.4}$& $\Delta k_{2}$ & $\Delta\Lambda_{1.4}$ \\
			&&&(km)&   & &(km)&& \\
			\cline{2-9}
			&$P_{\star}$&0&  13.166 &  0.0983 &685.4&$-$0.001&$-$0.0045&$-$33.1\\
			\cline{2-9}
			&\multirow{3}{*}{$P_{\mathrm{I}}$}
			& 0&  13.166 &  0.0997 &695.7&$-$0.001&$-$0.0031&$-$22.8\\
			&&1&  13.166 &  0.1020 &711.2&$-$0.001&$-$0.0008&$-$7.3\\
			&&2&  13.166 &  0.1025 &714.7&$-$0.001&$-$0.0003&$-$3.8\\
			\cline{2-9}
			&\multirow{3}{*}{$P_{\mathrm{II}}$}
			& 0&  13.166 &  0.1004 &700.6&$-$0.001&$-$0.0024&$-$17.9\\
			&&1&  13.166 &  0.1024 &714.0&$-$0.001&$-$0.0004&$-$4.5\\
			&&2&  13.166 &  0.1027 &716.5&$-$0.001&$-$0.0001&$-$2.0\\
			\cline{2-9}
			&\multirow{3}{*}{$P_{\mathrm{III}}$}
			& 0&  13.166 &  0.1009 &704.0&$-$0.001&$-$0.0019&$-$14.5\\
			&&1&  13.166 &  0.1025 &715.1&$-$0.001&$-$0.0003&$-$3.4\\
			&&2&  13.166 &  0.1028 &717.2&$-$0.001&  0 &$-$1.3\\
			\cline{2-9}
			&\multirow{3}{*}{$P_{\triangle}$}
			& 0&  13.166 &  0.1014 &707.5&$-$0.001&$-$0.0014&$-$11.0\\
			&&1&  13.166 &  0.1028 &717.3&$-$0.001&  0 &$-$1.2\\
			&&2&  13.166 &  0.1032 &719.6&$-$0.001&$+$0.0004&$+$1.1\\
			\hline\hline
			\multirow{8}{*}{\makecell[c]{Crossover connection}}
			&$P_C$ & $\Delta P$ &  $R_{1.4}$  & $k_2$ & $\Lambda_{1.4}$&$\Delta R_{1.4}$& $\Delta k_{2}$
			&$\Delta\Lambda_{1.4}$ \\
			&(MeV$/$fm$^{3}$)& (MeV$/$fm$^{3}$) &(km)&   & &(km)&& \\
			\cline{2-9}
			& \multirow{3}{*}{0.25}
			&  0.002 &  13.143 &  0.1037 &716.9&$-$0.001&$-$0.0002&$-$2.9\\
			&& 0.005 &  13.142 &  0.1037 &716.7&$-$0.002&$-$0.0002&$-$3.1\\
			&& 0.010 &  13.141 &  0.1037 &717.0&$-$0.003&$-$0.0002&$-$2.8\\
			\cline{2-9}
			& \multirow{3}{*}{0.30}
			&  0.002 &  13.157 &  0.1033 &718.2&$-$0.001&$-$0.0001&$-$1.9\\
			&& 0.005 &  13.157 &  0.1032 &717.3&$-$0.001&$-$0.0002&$-$2.8\\
			&& 0.010 &  13.157 &  0.1032 &717.6&$-$0.001&$-$0.0002&$-$2.5\\
			\hline\hline
			\multirow{7}{*}{\makecell[c]{Segmented method}}
			&\multicolumn{2}{c}{$P_S$}&  $R_{1.4}$  & $k_2$ & $\Lambda_{1.4}$&$\Delta R_{1.4}$& $\Delta k_{2}$
			&$\Delta\Lambda_{1.4}$ \\
			&\multicolumn{2}{c}{(MeV$/$fm$^{3}$)}&(km)&   & &(km)&& \\
			\cline{2-9}
			&\multicolumn{2}{c}{    0.20    }& 13.127 & 0.1041 &716.2&\multicolumn{3}{c}{\multirow{5}{*}{\diagbox[dir=NW]{}}}\\
			&\multicolumn{2}{c}{    0.25    }& 13.144 & 0.1039 &719.8\\
			&\multicolumn{2}{c}{    0.30    }& 13.158 & 0.1034 &720.1\\
			&\multicolumn{2}{c}{    0.35    }& 13.165 & 0.1031 &719.9\\
			&\multicolumn{2}{c}{$P_{\star}$}& 13.167 & 0.1028 &718.5\\
			\hline\hline
		\end{tabular}
		\label{ti14}
	\end{center}
\end{table*}

\begin{table*}[htbp]
	\begin{center}
		\caption{The same as Table~\ref{ti14} but for 
        the IUFSU (core) + TM1 (crust) combination.}
\begin{tabular}{c|cccccccccccccccccccccccccc}
			\hline\hline
			EOS& \multicolumn{8}{c}{\textbf{IUFSU (core) + TM1 (crust)}}\\
			\hline\hline
			\multirow{15}{*}{\makecell[c]{Direct connection}}
			&$P_D$&$n$&  $R_{1.4}$  & $k_2$ & $\Lambda_{1.4}$&$\Delta R_{1.4}$& $\Delta k_{2}$ & $\Delta\Lambda_{1.4}$ \\
			&&&(km)&   & &(km)&& \\
			\cline{2-9}
			&$P_{\star}$&0&  12.441 &  0.0757 &399.0&$-$0.002&$-$0.0194&$-$101.8\\
			\cline{2-9}
			&\multirow{3}{*}{$P_{\mathrm{I}}$}
			& 0&  12.443 &  0.0760 &401.1& 0 &$-$0.0191&$-$99.7\\
			&&1&  12.443 &  0.0782 &412.8& 0 &$-$0.0169&$-$88.0\\
			&&2&  12.443 &  0.0836 &440.8& 0 &$-$0.0115&$-$60.0\\
			\cline{2-9}
			&\multirow{3}{*}{$P_{\mathrm{II}}$}
			& 0&  12.445 &  0.0763 &403.0&$+$0.002&$-$0.0188&$-$97.8\\
			&&1&  12.444 &  0.0794 &419.1&$+$0.001&$-$0.0157&$-$81.7\\
			&&2&  12.444 &  0.0841 &444.0&$+$0.001&$-$0.0110&$-$56.8\\
			\cline{2-9}
			&\multirow{3}{*}{$P_{\mathrm{III}}$}
			& 0&  12.447 &  0.0766 &404.7&$+$0.004&$-$0.0185&$-$96.1\\
			&&1&  12.446 &  0.0801 &422.9&$+$0.003&$-$0.0150&$-$77.9\\
			&&2&  12.446 &  0.0842 &444.8&$+$0.003&$-$0.0109&$-$56.0\\
			\cline{2-9}
			&\multirow{3}{*}{$P_{\triangle}$}
			& 0&  12.448 &  0.0769 &406.3&$+$0.005&$-$0.0182&$-$94.5\\
			&&1&  12.447 &  0.0803 &424.5&$+$0.004&$-$0.0148&$-$76.3\\
			&&2&  12.446 &  0.0840 &444.0&$+$0.003&$-$0.0111&$-$56.8\\
			\hline\hline
			\multirow{8}{*}{\makecell[c]{Crossover connection}}
			&$P_C$ & $\Delta P$ &  $R_{1.4}$  & $k_2$ & $\Lambda_{1.4}$&$\Delta R_{1.4}$& $\Delta k_{2}$
			&$\Delta\Lambda_{1.4}$ \\
			&(MeV$/$fm$^{3}$)& (MeV$/$fm$^{3}$) &(km)&   & &(km)&& \\
			\cline{2-9}
			& \multirow{3}{*}{0.25}
			&  0.002 &  12.412 &  0.0959 &499.9&$-$0.002&$-$0.0007&$+$0.2\\
			&& 0.005 &  12.412 &  0.0959 &499.9&$-$0.002&$-$0.0007&$+$0.2\\
			&& 0.010 &  12.412 &  0.0960 &500.4&$-$0.002&$-$0.0006&$+$0.7\\
			\cline{2-9}
			& \multirow{3}{*}{0.30}
			&  0.002 &  12.428 &  0.0953 &500.1&$-$0.002&$-$0.0003&$-$0.5\\
			&& 0.005 &  12.428 &  0.0953 &500.2&$-$0.002&$-$0.0003&$-$0.4\\
			&& 0.010 &  12.428 &  0.0954 &500.8&$-$0.002&$-$0.0002&$+$0.2\\
			\hline\hline
			\multirow{6}{*}{\makecell[c]{Segmented method}}
			&\multicolumn{2}{c}{$P_S$}&  $R_{1.4}$  & $k_2$ & $\Lambda_{1.4}$&$\Delta R_{1.4}$& $\Delta k_{2}$
			&$\Delta\Lambda_{1.4}$ \\
			&\multicolumn{2}{c}{(MeV$/$fm$^{3}$)}&(km)&   & &(km)&& \\
			\cline{2-9}
			&\multicolumn{2}{c}{    0.20    }& 12.399 & 0.0966 &499.7&\multicolumn{3}{c}{\multirow{4}{*}{\diagbox[dir=NW]{}}}\\
			&\multicolumn{2}{c}{    0.25    }& 12.414 & 0.0961 &500.2\\
			&\multicolumn{2}{c}{    0.30    }& 12.430 & 0.0956 &500.6\\
			&\multicolumn{2}{c}{ $P_{\star}$}& 12.443 & 0.0951 &500.8\\
			\hline\hline
		\end{tabular}
		\label{IUFSU-1.4}
	\end{center}
\end{table*}

\begin{table*}[htbp]
	\begin{center}
		\caption{The same as Table~\ref{ti14} but for 
                the TM1e (core) + TM1 (crust) combination. }
		\begin{tabular}{c|cccccccccccccccccccccccccc}
			\hline\hline
			EOS& \multicolumn{8}{c}{\textbf{TM1e (core) + TM1 (crust)}}\\
			\hline\hline
			\multirow{15}{*}{\makecell[c]{Direct connection}}
			&$P_D$&$n$&  $R_{1.4}$  & $k_2$ & $\Lambda_{1.4}$&$\Delta R_{1.4}$& $\Delta k_{2}$ & $\Delta\Lambda_{1.4}$ \\
			&&&(km)&   & &(km)&& \\
			\cline{2-9}
			&$P_{\star}$&0&  12.935 &  0.0796 &509.0& 0 &$-$0.0259&$-$165.8\\
			\cline{2-9}
			&\multirow{3}{*}{$P_{\mathrm{I}}$}
			& 0&  12.936 &  0.0800 &511.9&$+$0.001&$-$0.0255&$-$162.9\\
			&&1&  12.936 &  0.0827 &529.1&$+$0.001&$-$0.0228&$-$145.7\\
			&&2&  12.936 &  0.0897 &574.1&$+$0.001&$-$0.0158&$-$100.7\\
			\cline{2-9}
			&\multirow{3}{*}{$P_{\mathrm{II}}$}
			& 0&  12.938 &  0.0803 &513.9&$+$0.003&$-$0.0252&$-$160.9\\
			&&1&  12.937 &  0.0840 &537.5&$+$0.002&$-$0.0215&$-$137.3\\
			&&2&  12.937 &  0.0903 &578.0&$+$0.002&$-$0.0152&$-$96.8\\
			\cline{2-9}
			&\multirow{3}{*}{$P_{\mathrm{III}}$}
			& 0&  12.940 &  0.0804 &515.2&$+$0.005&$-$0.0251&$-$159.6\\
			&&1&  12.938 &  0.0845 &541.3&$+$0.003&$-$0.0210&$-$133.5\\
			&&2&  12.938 &  0.0902 &577.7&$+$0.003&$-$0.0153&$-$97.1\\
			\cline{2-9}
			&\multirow{3}{*}{$P_{\triangle}$}
			&0 &  12.942 &  0.0814 &522.1&$+$0.007&$-$0.0241&$-$152.7\\
			&&1&  12.940 &  0.0858 &549.7&$+$0.005&$-$0.0197&$-$125.1\\
			&&2&  12.939 &  0.0905 &580.1&$+$0.004&$-$0.0150&$-$94.7\\
			\hline\hline
			\multirow{8}{*}{\makecell[c]{Crossover connection}}
			&$P_C$ & $\Delta P$ &  $R_{1.4}$  & $k_2$ & $\Lambda_{1.4}$&$\Delta R_{1.4}$& $\Delta k_{2}$
			&$\Delta\Lambda_{1.4}$ \\
			&(MeV$/$fm$^{3}$)& (MeV$/$fm$^{3}$) &(km)&   & &(km)&& \\
			\cline{2-9}
			& \multirow{3}{*}{0.25}
			&  0.002 &  12.899 &  0.1065 &672.1&$-$0.003&$-$0.0003&$-$1.8\\
			&& 0.005 &  12.900 &  0.1065 &672.3&$-$0.002&$-$0.0003&$-$1.6\\
			&& 0.010 &  12.900 &  0.1066 &672.7&$-$0.002&$-$0.0002&$-$1.2\\
			\cline{2-9}
			& \multirow{3}{*}{0.30}
			&  0.002 &  12.918 &  0.1058 &672.5&$-$0.002&$-$0.0003&$-$2.1\\
			&& 0.005 &  12.919 &  0.1059 &672.9&$-$0.001&$-$0.0002&$-$1.7\\
			&& 0.010 &  12.919 &  0.1059 &673.4&$-$0.001&$-$0.0002&$-$1.2\\
			\hline\hline
			\multirow{6}{*}{\makecell[c]{Segmented method}}
			&\multicolumn{2}{c}{$P_S$}&  $R_{1.4}$  & $k_2$ & $\Lambda_{1.4}$&$\Delta R_{1.4}$& $\Delta k_{2}$
			&$\Delta\Lambda_{1.4}$ \\
			&\multicolumn{2}{c}{(MeV$/$fm$^{3}$)}&(km)&   & &(km)&& \\
			\cline{2-9}
			&\multicolumn{2}{c}{    0.20    }& 12.884 & 0.1073 &672.4&\multicolumn{3}{c}{\multirow{4}{*}{\diagbox[dir=NW]{}}}\\
			&\multicolumn{2}{c}{    0.25    }& 12.902 & 0.1068 &673.9\\
			&\multicolumn{2}{c}{    0.30    }& 12.920 & 0.1061 &674.6\\
			&\multicolumn{2}{c}{$P_{\star}$}& 12.935 & 0.1055 &674.8\\
			\hline\hline
		\end{tabular}
		\label{TM1e-1.4}
	\end{center}
\end{table*}


In Fig.~\ref{pwtidal}, we show the tidal Love number $k_2$ (top panels) and the 
dimensionless tidal deformability $\Lambda$ (middle panels) as a function of the neutron-star 
mass $M$. These results are obtained using the segmented method. 
The $k_2$ values corresponding to different $P_S$ exhibit slight variations for massive stars. As the neutron-star mass decreases, the differences in $k_2$ among various $P_S$ increase and reach a maximum 
value at $M\approx$ 0.8$\Msun$. The insets provide more details for canonical neutron stars with masses 
around 1.4$\Msun$. It is shown that the increase in $k_2$ is negatively correlated with the increase 
in $P_S$. This implies that an earlier crust-core transition results in a smaller $k_2$, 
which is consistent with the behavior observed in Fig.~\ref{crtidal}. 
For the tidal deformability $\Lambda$, there is no significant difference among various 
values of $P_S$. According to Eq.~\eqref{eq:lambda}, for a neutron star with a certain mass, 
both the love number $k_2$ and the radius $R$ affect its tidal deformability $\Lambda$.
With increasing $P_S$, the decrease in $k_2$ can mostly offset the increase in the radius $R$, 
and as a result, no obvious dependence on $P_S$ is observed in $\Lambda$.
In Fig.~\ref{pwtidal}(a,b,c-3), we show the $y(r)$ profiles for a 1.4$\Msun$ 
neutron star obtained using the segmented method.
The $y(r)$ curves for the three EOS combinations exhibit a similar trend.
All curves are indistinguishable in the core region. 
When the crust-core transition occurs at $r_S$ (i.e., $P|_{r=r_S}=P_S$), a finite drop in the $y(r)$ 
profile appears and is indicated by a vertical dotted line, as described in Eq.~\eqref{eq:sy}. 
The dependence on $P_S$ can be observed in the insets. As $r$ increases, the differences in $y(r)$ 
among various $P_S$ gradually decrease, and ultimately reach almost identical values of $y_R$. 
Compared to the direct connection procedure, the uncertainty of $\Lambda$ in the segmented method 
is relatively small. This is because the fluctuation of the sound speed around
the crust-core transition does not affect the results in the segmented method, 
whereas it introduces more uncertainties in the tidal deformability for the direct connection procedure.
Furthermore, when constructing nonunified EOS, the crust-core transition point cannot be accurately 
determined. This means that the choice of $P_S$ involves a small degree of arbitrariness.
Therefore, the values of $P_S$ inevitably introduce some uncertainty into the predictions of neutron 
stars. In particular, the tidal deformability $\Lambda$ is relatively sensitive to the crust-core connection. 
Our results show that the segmented method yields significantly smaller uncertainties 
than the direct-connection procedure.

We present several properties of a 1.4$\Msun$ neutron star 
in Tables~\ref{ti14},~\ref{IUFSU-1.4},~\ref{TM1e-1.4} with three combinations of the crust 
and core EOSs, respectively. The results are obtained using the direct connection procedure, 
crossover connection procedure, and segmented method.
When using nonunified EOSs constructed by the direct connection, it is found that the radius of 
a 1.4$\Msun$ neutron star $R_{1.4}$ is insensitive to the connection details characterized by 
the parameters $P_D$ and $n$. 
In contrast, the tidal deformability $\Lambda_{1.4}$ 
is rather sensitive to variations of $P_D$ and $n$. Particularly, the results 
of $\Delta\Lambda_{1.4}$ with $n = 0$ are much larger than those in other cases.
This is because no additional data points are added to the connection segment to 
smooth the EOS in the case of $n = 0$. 
Increasing $n$ can reduce the differences in tidal deformabilities to a limited extent. 
For the combinations of IUFSU (core) + TM1 (crust) and TM1e (core) + TM1 (crust), 
even with $n=2$, the resulting values of $\Lambda_{1.4}$ still differ markedly from those 
of the segmented method.
The results obtained using the EOSs constructed by the crossover connection procedure are 
close to those obtained using the segmented method. The resulting values of $\Lambda_{1.4}$ 
are not very sensitive to the changes of $P_C$ and $\Delta P$ in the crossover connection procedure. 
Compared to the direct and crossover connection procedures, the segmented method provides 
consistent estimates for $\Lambda_{1.4}$, and the results are not very sensitive to the 
changes of $P_S$ in the segmented method.

\section{Summary}
\label{sec:4}

In this work, we conducted a detailed study on the impact of the crust-core connection procedure 
when constructing a nonunified EOS for neutron stars, where the crust and core segments are 
obtained within different models. Generally, a finite energy density discontinuity is often 
present in a nonunified EOS around the crust-core transition, which may induce uncertainties 
in the prediction of the tidal deformability. We considered three types of connection procedures 
to treat this discontinuity.
The first one, i.e., the direct connection procedure, adopts the Newton interpolation 
method to directly connect the inner crust and core segments. 
The second one, i.e., the crossover connection procedure, employs the regularized 
calculation to generate a crossover EOS between the inner crust and core segments.
The third one, i.e., the segmented method, solves the TOV equation separately
inside the crust and core regions, while appropriate matching conditions are imposed 
at the crust-core interface. 

Our results indicate that the mass-radius relations of neutron stars are almost unaffected 
by the details of the connection procedure. However, tidal deformabilities obtained using 
different connection procedures may have some uncertainties. 
For a canonical 1.4$\Msun$ neutron star, uncertainties 
in the tidal deformability $\Lambda_{1.4}$ from different connection procedures can exceed 20\%.
This is because the fluctuation of the sound speed around the crust-core transition
can significantly affect the results of tidal deformabilities.
Among the three types of connection procedures considered in the present work,
the direct connection yields significantly larger uncertainties in $\Lambda_{1.4}$ than 
the crossover connection and segmented method.
To avoid the influence caused by the fluctuation of the sound speed, we employ the segmented method,
in which the TOV equation is solved in segments, with appropriate matching conditions 
at the crust-core interface. The advantage of the segmented method is that no additional
modifications are made to the original crust and core EOS data. 
As a result, the uncertainties in the segmented method are much smaller than 
those in the direct connection procedure. On the other hand, the crossover connection introduces 
an additional EOS segment, which is generated using a regularized calculation and bridges the inner 
crust and core segments. It is found that both the crossover connection and segmented method 
can provide stable results for the tidal deformability. We emphasize that when employing a 
nonunified EOS to study neutron-star properties, especially the tidal deformability, 
special attention should be paid to the crust-core connection procedure.

\section{Acknowledgments}

This work was partially supported by the National Natural Science Foundation of
China under Grant Nos. 12175109 and 12475149, and by Guangdong Basic and Applied
 Basic Research Foundation under Grant No. 2024A1515010911.

\bibliographystyle{apsrev4-1}
\bibliography{refs-pang}

@article{abbo16,
  title = {Observation of Gravitational Waves from a Binary Black Hole Merger},
  author = {Abbott, B. P. and Abbott, R. and Abbott, T. D. and Abernathy, M. R. and Acernese, F. and Ackley, K. and Adams, C. and Adams, T. and Addesso, P. and others},
  collaboration = {LIGO Scientific Collaboration and Virgo Collaboration},
  journal = {Physical Review Letters},
  volume = {116},
  issue = {6},
  pages = {061102},
  numpages = {16},
  year = {2016},
  month = {Feb},
  publisher = {American Physical Society},
  doi = {10.1103/PhysRevLett.116.061102},
  url = {https://link.aps.org/doi/10.1103/PhysRevLett.116.061102}
}

@article{abbo17,
  title={GW170817: observation of gravitational waves from a binary neutron star
  inspiral},
  author={Abbott, Benjamin P and Abbott, Rich and Abbott, TD and Acernese, Fausto and
  Ackley, Kendall and Adams, Carl and Adams, Thomas and Addesso, Paolo and Adhikari, RX
  and Adya, Vaishali B and others},
      journal = {\prl},
         year = 2017,
        month = oct,
       volume = {119},
       number = {16},
        pages = {161101},
          doi = {10.1103/PhysRevLett.119.161101},
          url = {https://doi.org/10.1103/PhysRevLett.119.161101}
}

@article{abbo18,
  title={GW170817: Measurements of neutron star radii and equation of state},
  author={Abbott, Benjamin P and Abbott, Richard and Abbott, TD and Acernese, F and
  Ackley, K and Adams, C and Adams, T and Addesso, P and Adhikari, Rana X and Adya,
  Vaishali B and others},
  journal={\prl},
  volume = {121},
  issue = {16},
  pages = {161101},
  numpages = {16},
  year = {2018},
  month = {Oct},
  publisher = {American Physical Society},
  doi = {10.1103/PhysRevLett.121.161101},
  url = {https://link.aps.org/doi/10.1103/PhysRevLett.121.161101}
}

@article{alfo17,
doi = {10.3847/1538-4357/aa8509},
url = {https://dx.doi.org/10.3847/1538-4357/aa8509},
year = {2017},
month = {sep},
publisher = {\aps},
volume = {847},
number = {2},
pages = {109},
author = {Mark G. Alford and Steven P. Harris and Pratik S. Sachdeva},
title = {On the Stability of Strange Dwarf Hybrid Stars},
journal = {\apj},
}

@article{anna20,
  title={Evidence for quark-matter cores in massive neutron stars},
  author={Annala, Eemeli and Gorda, Tyler and Kurkela, Aleksi and N{\"a}ttil{\"a}, Joonas and Vuorinen, Aleksi},
  journal={Nature Physics},
  volume={16},
  number={9},
  pages={907--910},
  year={2020},
  publisher={Nature Publishing Group UK London},
  doi = {10.1038/s41567-020-0914-9},
}

@article{anto13,
title = {A Massive Pulsar in a Compact Relativistic Binary},
author = {John Antoniadis  and Paulo C. C. Freire  and Norbert Wex  and others},
journal = {Science},
volume = {340},
number = {6131},
pages = {1233232},
year = {2013},
doi = {10.1126/science.1233232},
URL = {https://www.science.org/doi/abs/10.1126/science.1233232}
}

@article{arzo18,
title = {The NANOGrav 11-year Data Set: High-precision Timing of 45 Millisecond Pulsars},
author = {Zaven Arzoumanian and Adam Brazier and Sarah Burke-Spolaor and others},
journal = {\apjs},
year = {2018},
month = {apr},
publisher = {The American Astronomical Society},
volume = {235},
number = {2},
pages = {37},
doi = {10.3847/1538-4365/aab5b0},
url = {https://dx.doi.org/10.3847/1538-4365/aab5b0},
}

@article{asce24,
title = {Neutron-star measurements in the multi-messenger Era},
author = {Stefano Ascenzi and Vanessa Graber and Nanda Rea},
journal = {Astroparticle Physics},
volume = {158},
pages = {102935},
year = {2024},
issn = {0927-6505},
doi = {https://doi.org/10.1016/j.astropartphys.2024.102935},
url = {https://www.sciencedirect.com/science/article/pii/S0927650524000124}
}

@article{avan08,
  title = {Warm and cold pasta phase in relativistic mean field theory},
  author = {Avancini, S. S. and Menezes, D. P. and Alloy, M. D. and Marinelli, J. R. and Moraes, M. M. W. and Provid\^encia, C.},
  journal = {\prc},
  volume = {78},
  issue = {1},
  pages = {015802},
  numpages = {12},
  year = {2008},
  month = {Jul},
  publisher = {American Physical Society},
  doi = {10.1103/PhysRevC.78.015802},
  url = {https://link.aps.org/doi/10.1103/PhysRevC.78.015802}
}

@article{bao14a,
  title = {Influence of the symmetry energy on nuclear pasta in neutron star crusts},
  author = {Bao, Shi Shao and Shen, Hong},
  journal = {\prc},
  volume = {89},
  issue = {4},
  pages = {045807},
  year = {2014},
  month = {Apr},
  publisher = {American Physical Society},
  doi = {10.1103/PhysRevC.89.045807},
  url = {https://link.aps.org/doi/10.1103/PhysRevC.89.045807}
}

@article{bao14b,
  title = {Effects of the symmetry energy on properties of neutron star crusts near the
  neutron drip density},
  author = {Bao, Shi Shao and Hu, Jin Niu and Zhang, Zhao Wen and Shen, Hong},
  journal = {\prc},
  volume = {90},
  issue = {4},
  pages = {045802},
  year = {2014},
  month = {Oct},
  publisher = {American Physical Society},
  doi = {10.1103/PhysRevC.90.045802},
  url = {https://link.aps.org/doi/10.1103/PhysRevC.90.045802}
}

@article{bao15,
  title={Impact of the symmetry energy on nuclear pasta phases and crust-core transition
  in neutron stars},
  author={Bao, Shi Shao and Shen, Hong},
  journal = {\prc},
  volume = {91},
  issue = {1},
  pages = {015807},
  numpages = {10},
  year = {2015},
  month = {Jan},
  publisher = {American Physical Society},
  doi = {10.1103/PhysRevC.91.015807},
  url = {https://link.aps.org/doi/10.1103/PhysRevC.91.015807}
}

@article{baym71,
title = {Neutron star matter},
author = {Gordon Baym and Hans A. Bethe and Christopher J Pethick},
journal = {Nucl. Phys. A},
volume = {175},
number = {2},
pages = {225-271},
year = {1971},
issn = {0375-9474},
doi = {https://doi.org/10.1016/0375-9474(71)90281-8},
url = {https://www.sciencedirect.com/science/article/pii/0375947471902818}
}

@article{cham08,
       author = {{Chamel}, Nicolas and {Haensel}, Pawel},
        title = "{Physics of Neutron Star Crusts}",
      journal = {Living Rev. Relativ.},
         year = 2008,
        month = dec,
       volume = {11},
       number = {1},
          eid = {10},
        pages = {10},
          doi = {10.12942/lrr-2008-10}
}

@article{char14,
  title = {Massive neutron stars with antikaon condensates in a density-dependent hadron field theory},
  author = {Char, Prasanta and Banik, Sarmistha},
  journal = {Physical Review C},
  volume = {90},
  issue = {1},
  pages = {015801},
  numpages = {9},
  year = {2014},
  month = {Jul},
  publisher = {American Physical Society},
  doi = {10.1103/PhysRevC.90.015801},
  url = {https://link.aps.org/doi/10.1103/PhysRevC.90.015801}
}

@article{chat20,
    author = "Chatziioannou, Katerina",
    title = "{Neutron star tidal deformability and equation of state constraints}",
    journal = "General Relativity and Gravitation",
    volume = "52",
    number = "11",
    pages = "109",
    year = "2020",
    doi = "10.1007/s10714-020-02754-3",
}

@article{chat24,
  title = {Neutron stars and the dense matter equation of state},
  author = {Chatziioannou, Katerina and Cromartie, H. Thankful and Gandolfi, Stefano and Tews, Ingo and Radice, David and Steiner, Andrew W. and Watts, Anna L.},
  journal = {Rev. Mod. Phys.},
  volume = {97},
  issue = {4},
  pages = {045007},
  numpages = {49},
  year = {2025},
  month = {Dec},
  publisher = {American Physical Society},
  doi = {10.1103/ymsq-cfcw},
  url = {https://link.aps.org/doi/10.1103/ymsq-cfcw}
}

@article{douc01,
    author = "Douchin, F. and Haensel, P.",
    title = "{A unified equation of state of dense matter and neutron star structure}",
    journal = "Astronomy \& astrophysics",
    volume = "380",
    pages = "151",
    year = "2001",
    doi = "10.1051/0004-6361:20011402",
}

@article{dutr14,
  title={Relativistic mean-field hadronic models under nuclear matter constraints},
  author={Dutra, M and Louren{\c{c}}o, O and Avancini, SS and Carlson, B V and Delfino, A
  and Menezes, DP and Provid{\^e}ncia, C and Typel, S and Stone, JR},
  journal={\prc},
  volume={90},
  number={5},
  pages={055203},
  year={2014},
  publisher = {American Physical Society},
  doi = {10.1103/PhysRevC.90.055203},
  url = {https://link.aps.org/doi/10.1103/PhysRevC.90.055203}
}

@article{fant13,
  title={Neutron star properties with unified equations of state of dense matter},
  author={Fantina, AF and Chamel, N and Pearson, JM and Goriely, S},
  journal={Astronomy \& astrophysics},
  volume={559},
  pages={A128},
  year={2013},
  publisher={EDP Sciences},
  doi = {10.1051/0004-6361/201321884}
}

@article{fatt18,
  title = {Neutron Skins and Neutron Stars in the Multimessenger Era},
  author = {Fattoyev, F. J. and Piekarewicz, J. and Horowitz, C. J.},
  journal = {\prl},
  volume = {120},
  issue = {17},
  pages = {172702},
  numpages = {6},
  year = {2018},
  month = {Apr},
  publisher = {American Physical Society},
  doi = {10.1103/PhysRevLett.120.172702},
  url = {https://link.aps.org/doi/10.1103/PhysRevLett.120.172702}
}

@article{fatt20,
  title={GW190814: Impact of a 2.6 solar mass neutron star on the nucleonic equations of
  state},
  author={Fattoyev, FJ and Horowitz, CJ and Piekarewicz, J and Reed, Brendan},
  journal={\prc},
  volume={102},
  number={6},
  pages={065805},
  year={2020},
  month = {Dec},
  publisher = {American Physical Society},
  doi = {10.1103/PhysRevC.102.065805},
  url = {https://link.aps.org/doi/10.1103/PhysRevC.102.065805}
}

@article{fons21,
title = {Refined Mass and Geometric Measurements of the High-mass PSR J0740+6620},
author = {E. Fonseca and H. T. Cromartie and T. T. Pennucci and others},
journal = {\apjl},
year = {2021},
month = {jul},
publisher = {The American Astronomical Society},
volume = {915},
number = {1},
pages = {L12},
doi = {10.3847/2041-8213/ac03b8},
url = {https://dx.doi.org/10.3847/2041-8213/ac03b8},
}

@article{fort16,
  title = {Neutron star radii and crusts: Uncertainties and unified equations of state},
  author = {Fortin, M. and Provid\^encia, C. and Raduta, Ad. R. and Gulminelli, F. and
  Zdunik, J. L. and Haensel, P. and Bejger, M.},
  journal = {\prc},
  volume = {94},
  issue = {3},
  pages = {035804},
  numpages = {21},
  year = {2016},
  month = {Sep},
  publisher = {American Physical Society},
  doi = {10.1103/PhysRevC.94.035804},
  url = {https://link.aps.org/doi/10.1103/PhysRevC.94.035804}
}

@article{gasc00,
  title={Polynomial interpolation in several variables},
  author={Gasca, Mariano and Sauer, Thomas},
  journal={Advances in Computational Mathematics},
  volume={12},
  pages={377--410},
  year={2000},
  publisher={Springer},
  doi={10.1023/A:1018981505752},
  url={https://doi.org/10.1023/A:1018981505752}
}

@article{gulm15,
  title = {Unified treatment of subsaturation stellar matter at zero and finite temperature},
  author = {Gulminelli, F. and Raduta, Ad. R.},
  journal = {Physical Review C},
  volume = {92},
  issue = {5},
  pages = {055803},
  numpages = {26},
  year = {2015},
  month = {Nov},
  publisher = {American Physical Society},
  doi = {10.1103/PhysRevC.92.055803},
  url = {https://link.aps.org/doi/10.1103/PhysRevC.92.055803}
}

@article{han19,
  title = {Tidal deformability with sharp phase transitions in binary neutron stars},
  author = {Han, Sophia and Steiner, Andrew W.},
  journal = {Physical Review D},
  volume = {99},
  issue = {8},
  pages = {083014},
  numpages = {21},
  year = {2019},
  month = {Apr},
  publisher = {American Physical Society},
  doi = {10.1103/PhysRevD.99.083014},
  url = {https://link.aps.org/doi/10.1103/PhysRevD.99.083014}
}

@article{hind08,
author = {Hinderer, Tanja},
title = {Tidal Love Numbers of Neutron Stars},
journal = {The Astrophysical Journal},
year = {2008},
month = {apr},
publisher = {},
volume = {677},
number = {2},
pages = {1216},
doi = {10.1086/533487},
url = {https://doi.org/10.1086/533487}
}

@article{huang22,
  title={The Hadron-quark Crossover in Neutron Star within Gaussian Process Regression
  Method},
  author={Huang, Kai Xuan and Hu, Jin Niu and Zhang, Ying and Shen, Hong},
  journal={\apj},
  volume={935},
  number={2},
  pages={88},
  year={2022},
  doi = {10.3847/1538-4357/ac7f3c},
  url = {https://dx.doi.org/10.3847/1538-4357/ac7f3c},
  publisher={IOP Publishing}
}

@article{huang22b,
    author = {Huang, Kaixuan and Hu, Jinniu and Zhang, Ying and Shen, Hong},
    title = {Investigation on the Hyperonic Star in Relativistic Mean-field Model},
    journal = {Nuclear Physics Review},
    volume = {39},
    number = {2},
    pages = {135--153},
    year = {2022},
    doi = {10.11804/NuclPhysRev.39.2022013}
}

@article{ji19,
  title = {Effects of nuclear symmetry energy and equation of state on neutron star
  properties},
  author = {Ji, Fan and Hu, Jin Niu and Bao, Shi Shao and Shen, Hong},
  journal = {\prc},
  volume = {100},
  issue = {4},
  pages = {045801},
  numpages = {11},
  year = {2019},
  month = {Oct},
  publisher = {American Physical Society},
  doi = {10.1103/PhysRevC.100.045801},
  url = {https://link.aps.org/doi/10.1103/PhysRevC.100.045801}
}

@article{ju21,
title = {Hadron-quark Pasta Phase in Massive Neutron Stars},
author = {Min Ju and Jin Niu Hu and Hong Shen},
journal = {\apj},
volume = {923},
number = {2},
pages = {250},
year = {2021},
publisher = {The American Astronomical Society},
doi = {10.3847/1538-4357/ac30dd},
url = {https://dx.doi.org/10.3847/1538-4357/ac30dd}
}

@Article{khun19,
AUTHOR = {Khunjua, Tamaz and Klimenko, Konstantin and Zhokhov, Roman},
TITLE = {Charged Pion Condensation in Dense Quark Matter: Nambu–Jona-Lasinio Model Study},
JOURNAL = {Symmetry},
VOLUME = {11},
YEAR = {2019},
NUMBER = {6},
ARTICLE-NUMBER = {778},
URL = {https://www.mdpi.com/2073-8994/11/6/778},
ISSN = {2073-8994},
DOI = {10.3390/sym11060778}
}

@article{koeh25,
  title = {From Existing and New Nuclear and Astrophysical Constraints to Stringent Limits on the Equation of State of Neutron-Rich Dense Matter},
  author = {Koehn, Hauke and Rose, Henrik and Pang, Peter T. H. and Somasundaram, Rahul and Reed, Brendan T. and Tews, Ingo and Abac, Adrian and Komoltsev, Oleg and Kunert, Nina and Kurkela, Aleksi and Coughlin, Michael W. and Healy, Brian F. and Dietrich, Tim},
  journal = {Physical Review X},
  volume = {15},
  issue = {2},
  pages = {021014},
  numpages = {55},
  year = {2025},
  month = {Apr},
  publisher = {American Physical Society},
  doi = {10.1103/PhysRevX.15.021014},
  url = {https://link.aps.org/doi/10.1103/PhysRevX.15.021014}
}

@article{kund23,
  title = {(Anti)kaon condensation in strongly magnetized dense matter},
  author = {Kundu, Debraj and Thapa, Vivek Baruah and Sinha, Monika},
  journal = {Physical Review C},
  volume = {107},
  issue = {3},
  pages = {035807},
  numpages = {14},
  year = {2023},
  month = {Mar},
  publisher = {American Physical Society},
  doi = {10.1103/PhysRevC.107.035807},
  url = {https://link.aps.org/doi/10.1103/PhysRevC.107.035807}
}

@article{land20,
  title = {Nonparametric constraints on neutron star matter with existing and upcoming gravitational wave and pulsar observations},
  author = {Landry, Philippe and Essick, Reed and Chatziioannou, Katerina},
  journal = {Physical Review D},
  volume = {101},
  issue = {12},
  pages = {123007},
  numpages = {21},
  year = {2020},
  month = {Jun},
  publisher = {American Physical Society},
  doi = {10.1103/PhysRevD.101.123007},
  url = {https://link.aps.org/doi/10.1103/PhysRevD.101.123007}
}

@article{latt16,
title = {The equation of state of hot, dense matter and neutron stars},
author = {James M. Lattimer and Madappa Prakash},
journal = {Phys. Rep.},
volume = {621},
pages = {127-164},
year = {2016},
note = {Memorial Volume in Honor of Gerald E. Brown},
issn = {0370-1573},
doi = {https://doi.org/10.1016/j.physrep.2015.12.005},
url = {https://www.sciencedirect.com/science/article/pii/S0370157315005396}
}

@Article{logo21,
AUTHOR = {Logoteta, Domenico},
TITLE = {Hyperons in Neutron Stars},
JOURNAL = {Universe},
VOLUME = {7},
YEAR = {2021},
NUMBER = {11},
ARTICLE-NUMBER = {408},
URL = {https://www.mdpi.com/2218-1997/7/11/408},
ISSN = {2218-1997},
DOI = {10.3390/universe7110408}
}

@article{mill19,
  title={PSR J0030+ 0451 mass and radius from NICER data and implications for the properties of neutron star matter},
  author={Miller, M. Coleman  and Lamb, Frederick K and Dittmann, A J and Bogdanov, Slavko and Arzoumanian, Zaven and Gendreau, Keith C and Guillot, S and Harding, AK and Ho, WCG and Lattimer, J M and others},
  journal={\apjl},
  volume={887},
  number={1},
  pages={L24},
  year = {2019},
  month = {dec},
  publisher = {The American Astronomical Society},
  doi = {10.3847/2041-8213/ab50c5},
  url = {https://dx.doi.org/10.3847/2041-8213/ab50c5},
}

@article{mill21,
  title={The radius of PSR J0740+ 6620 from NICER and XMM-Newton data},
  author={Miller, M. Coleman and Lamb, F K and Dittmann, A J and Bogdanov, S and Arzoumanian, Z and Gendreau, K C and Guillot, S and Ho, W C G and Lattimer, J M and Loewenstein, M and others},
  journal={\apjl},
  volume={918},
  number={2},
  pages={L28},
  year = {2021},
  month = {sep},
  publisher = {The American Astronomical Society},
  doi = {10.3847/2041-8213/ac089b},
  url = {https://dx.doi.org/10.3847/2041-8213/ac089b},
}

@article{miya13,
doi = {10.1088/0004-637X/777/1/4},
url = {https://doi.org/10.1088/0004-637X/777/1/4},
year = {2013},
month = {oct},
publisher = {The American Astronomical Society},
volume = {777},
number = {1},
pages = {4},
author = {Miyatsu, Tsuyoshi and Yamamuro, Sachiko and Nakazato, Ken'ichiro},
title = {A NEW EQUATION OF STATE FOR NEUTRON STAR MATTER WITH NUCLEI IN THE CRUST AND HYPERONS IN THE CORE},
journal = {\apj}
}

@article{most18,
  title = {New Constraints on Radii and Tidal Deformabilities of Neutron Stars from GW170817},
  author = {Most, Elias R. and Weih, Lukas R. and Rezzolla, Luciano and Schaffner-Bielich, J\"urgen},
  journal = {\prl},
  volume = {120},
  issue = {26},
  pages = {261103},
  numpages = {6},
  year = {2018},
  month = {Jun},
  publisher = {American Physical Society},
  doi = {10.1103/PhysRevLett.120.261103},
  url = {https://link.aps.org/doi/10.1103/PhysRevLett.120.261103}
}

@article{oert17,
  title = {Equations of state for supernovae and compact stars},
  author = {Oertel, M. and Hempel, M. and Kl\"ahn, T. and Typel, S.},
  journal = {Reviews of Modern Physics},
  volume = {89},
  issue = {1},
  pages = {015007},
  year = {2017},
  month = {Mar},
  publisher = {American Physical Society},
  doi = {10.1103/RevModPhys.89.015007},
  url = {https://link.aps.org/doi/10.1103/RevModPhys.89.015007}
}

@article{ohni09,
  title = {Possibility of an $s$-wave pion condensate in neutron stars reexamined},
  author = {Ohnishi, A. and Jido, D. and Sekihara, T. and Tsubakihara, K.},
  journal = {Physical Review C},
  volume = {80},
  issue = {3},
  pages = {038202},
  numpages = {4},
  year = {2009},
  month = {Sep},
  publisher = {American Physical Society},
  doi = {10.1103/PhysRevC.80.038202},
  url = {https://link.aps.org/doi/10.1103/PhysRevC.80.038202}
}

@article{oppe39,
  title={On massive neutron cores},
  author={Oppenheimer, J Robert and Volkoff, George M},
  journal={Physical Review},
  volume={55},
  number={4},
  pages={374},
  year={1939},
  doi = {10.1103/PhysRev.55.374},
  url = {https://link.aps.org/doi/10.1103/PhysRev.55.374},
  publisher={APS}
}

@article{post10,
  title = {Tidal Love numbers of neutron and self-bound quark stars},
  author = {Postnikov, Sergey and Prakash, Madappa and Lattimer, James M.},
  journal = {\prd},
  volume = {82},
  issue = {2},
  pages = {024016},
  numpages = {12},
  year = {2010},
  month = {Jul},
  publisher = {American Physical Society},
  doi = {10.1103/PhysRevD.82.024016},
  url = {https://link.aps.org/doi/10.1103/PhysRevD.82.024016}
}

@article{rave83,
  title = {Structure of Matter below Nuclear Saturation Density},
  author = {Ravenhall, D. G. and Pethick, C. J. and Wilson, J. R.},
  journal = {\prl},
  volume = {50},
  issue = {26},
  pages = {2066--2069},
  numpages = {0},
  year = {1983},
  month = {Jun},
  publisher = {American Physical Society},
  doi = {10.1103/PhysRevLett.50.2066},
  url = {https://link.aps.org/doi/10.1103/PhysRevLett.50.2066}
}

@article{rile19,
  title={A NICER view of PSR J0030+ 0451: Millisecond pulsar parameter estimation},
  author={Riley, Thomas E and Watts, Anna L and Bogdanov, Slavko and Ray, Paul S and Ludlam, Renee M and Guillot, Sebastien and Arzoumanian, Zaven and Baker, Charles L and Bilous, Anna V and Chakrabarty, Deepto and others},
  journal={\apjl},
  volume={887},
  number={1},
  pages={L21},
  year={2019},
  doi = {10.3847/2041-8213/ab481c},
  url = {https://dx.doi.org/10.3847/2041-8213/ab481c},
  publisher={IOP Publishing}
}

@article{rile21,
  title={A NICER view of the massive pulsar PSR J0740+ 6620 informed by radio timing and XMM-Newton spectroscopy},
  author={Riley, Thomas E and Watts, Anna L and Ray, Paul S and Bogdanov, Slavko and Guillot, Sebastien and Morsink, Sharon M and Bilous, Anna V and Arzoumanian, Zaven and Choudhury, Devarshi and Deneva, Julia S and others},
  journal={\apjl},
  volume={918},
  number={2},
  pages={L27},
  year={2021},
  doi = {10.3847/2041-8213/ac0a81},
  url = {https://dx.doi.org/10.3847/2041-8213/ac0a81},
  publisher={IOP Publishing}
}

@article{shen02,
  title = {Complete relativistic equation of state for neutron stars},
  author = {Shen, H.},
  journal = {Phys. Rev. C},
  volume = {65},
  issue = {3},
  pages = {035802},
  numpages = {7},
  year = {2002},
  month = {Feb},
  publisher = {American Physical Society},
  doi = {10.1103/PhysRevC.65.035802},
  url = {https://link.aps.org/doi/10.1103/PhysRevC.65.035802}
}

@article{shen20,
  title={Effects of symmetry energy on the equation of state for simulations of core-collapse supernovae and neutron-star mergers},
  author={Shen, Hong and Ji, Fan and Hu, Jin Niu and Sumiyoshi, Kohsuke},
  journal={\apj},
  volume={891},
  number={2},
  pages={148},
  year={2020},
  publisher = {The American Astronomical Society},
  doi = {10.3847/1538-4357/ab72fd},
  url = {https://dx.doi.org/10.3847/1538-4357/ab72fd},
}

@article{tu22,
title = {Effects of the phi Meson on the Properties of Hyperon Stars in the Density-dependent Relativistic Mean Field Model},
author = {Tu, Zhong-Hao and Zhou, Shan-Gui},
journal = {The Astrophysical Journal},
year = {2022},
month = {jan},
volume = {925},
number = {1},
pages = {16},
publisher = {The American Astronomical Society},
doi = {10.3847/1538-4357/ac3996},
url = {https://doi.org/10.3847/1538-4357/ac3996}
}

@article{webe05,
  title={Strange quark matter and compact stars},
  author={Weber, Fridolin},
  journal={Progress in Particle and Nuclear Physics},
  volume={54},
  number={1},
  pages={193--288},
  year={2005},
  doi = {https://doi.org/10.1016/j.ppnp.2004.07.001},
  url = {https://www.sciencedirect.com/science/article/pii/S0146641004001061},
  publisher={Elsevier}
}

@article{wu19,
  title = {Nuclear symmetry energy and hadron-quark mixed phase in neutron stars},
  author = {Wu, Xu Hao and Shen, Hong},
  journal = {\prc},
  volume = {99},
  issue = {6},
  pages = {065802},
  year = {2019},
  month = {Jun},
  publisher = {American Physical Society},
  doi = {10.1103/PhysRevC.99.065802},
  url = {https://link.aps.org/doi/10.1103/PhysRevC.99.065802}
}

@article{yue08,
  title = {Quark-meson coupling model for antikaon condensation in neutron star matter with strong magnetic fields},
  author = {Yue, P. and Shen, H.},
  journal = {Physical Review C},
  volume = {77},
  issue = {4},
  pages = {045804},
  numpages = {8},
  year = {2008},
  month = {Apr},
  publisher = {American Physical Society},
  doi = {10.1103/PhysRevC.77.045804},
  url = {https://link.aps.org/doi/10.1103/PhysRevC.77.045804}
}

@article{xia24,
  title = {Mixed phases of compact star matter in a unified mean-field approach},
  author = {Xia, Cheng-Jun and Maruyama, Toshiki and Yasutake, Nobutoshi and Tatsumi, Toshitaka},
  journal = {Physical Review D},
  volume = {110},
  issue = {11},
  pages = {114024},
  numpages = {14},
  year = {2024},
  month = {Dec},
  publisher = {American Physical Society},
  doi = {10.1103/PhysRevD.110.114024},
  url = {https://link.aps.org/doi/10.1103/PhysRevD.110.114024}
}

@article{yang08,
  title={Influence of the hadronic equation of state on the hadron-quark phase transition in neutron stars},
  author={Yang, F and Shen, H},
  journal={Physical Review C},
  volume = {77},
  issue = {2},
  pages = {025801},
  numpages = {8},
  year = {2008},
  month = {Feb},
  publisher = {American Physical Society},
  doi = {10.1103/PhysRevC.77.025801},
  url = {https://link.aps.org/doi/10.1103/PhysRevC.77.025801}
}

@article{piek2010,
  title = {Relativistic effective interaction for nuclei, giant resonances, and neutron stars},
  author = {Fattoyev, F. J. and Horowitz, C. J. and Piekarewicz, J. and Shen, G.},
  journal = {Phys. Rev. C},
  volume = {82},
  issue = {5},
  pages = {055803},
  numpages = {8},
  year = {2010},
  month = {Nov},
  publisher = {American Physical Society},
  doi = {10.1103/PhysRevC.82.055803},
  url = {https://link.aps.org/doi/10.1103/PhysRevC.82.055803}
}

\end{document}